\documentclass[%
 reprint,
 amsmath,amssymb,
 aps,
]{revtex4-2}

\usepackage{graphicx}
\usepackage{dcolumn}
\usepackage{bm}
\usepackage{hyperref}

\usepackage{amsmath,amsthm, amssymb} 
\usepackage{dsfont,yfonts}
\usepackage{verbatim}
\usepackage{graphics,graphicx}

\hypersetup{colorlinks=true,linkcolor=blue,citecolor=red,urlcolor=blue}

\usepackage{xparse}

\newcommand{\MeijerG}[7]{G \begin{smallmatrix} #1,\!\!&#2\\#3,\!\!&#4 \end{smallmatrix} \left( \begin{smallmatrix} #5 \\ #6 \end{smallmatrix} \middle\vert #7 \right) }

\begin{document}

\preprint{APS/123-QED}

\title{Geometry-free renormalization of directed networks: scale-invariance and reciprocity}

\author{Margherita Lalli}
\affiliation{Scuola Normale Superiore, Piazza dei Cavalieri, Pisa (Italy).}
\affiliation{ 
IMT School for Advanced Studies, Piazza San Francesco 19, 55100 Lucca (Italy)
}%
\author{Diego Garlaschelli}%
 \email{E-mail: diego.garlaschelli@imtlucca.it}
 \affiliation{ 
IMT School for Advanced Studies, Piazza San Francesco 19, 55100 Lucca (Italy)
}%
\affiliation{%
Lorentz Institute for Theoretical Physics, University of Leiden, Niels Bohrweg 2, 2333 CA Leiden (The Netherlands)
}%
\affiliation{INdAM-GNAMPA Istituto Nazionale di Alta Matematica (Italy)}

\date{\today}

\begin{abstract}
Recent research has tried to extend the concept of renormalization, which is naturally defined for geometric objects, to more general networks with arbitrary topology.
The current attempts do not naturally apply to directed networks, for instance because they are based on the identification of (necessarily symmetric) inter-node distances arising from geometric embeddings or on the definition of Hermitian Laplacian operators requiring symmetric adjacency matrices in spectral approaches.
Here we show that the Scale-Invariant Model, recently proposed as an approach to consistently model undirected networks at arbitrary (and possibly multi-scale) resolution levels, can be extended coherently to directed networks without the requirement of an embedding geometry or Laplacian structure. 
Moreover, it can account for nontrivial reciprocity, i.e. the empirically well-documented tendency of links to occur in mutual pairs more or less often than predicted by chance.
After deriving the renormalization rules for networks with arbitrary reciprocity, we consider various examples. In particular we propose the first multiscale model of the international trade network with nontrivial reciprocity and an annealed model where positive reciprocity emerges spontaneously from coarse-graining.
In the latter case, the renormalization process defines a group, not only a semigroup, and therefore allows to fine-grain networks to arbitrarily small scales.
These results strengthen the notion that network renormalization can be defined in a much more general way than required by geometric or spectral approaches, because it needs only node-specific (metric-free) features and can coexist with the asymmetry of directed networks. 
\end{abstract}

\maketitle

\section{Introduction}

Complex systems typically exhibit structural features coexisting over multiple characteristic scales~\cite{anderson1972more,adami2002complexity}.
This calls for robust multiscale representations and models capable of capturing properties at arbitrary levels of resolution, for both practical and theoretical reasons. 
The Renormalization Group (RG) is a powerful approach in statistical mechanics~\cite{kadanoff2000statistical,wilson1983renormalization, goldenfeld2018lectures} that identifies a repeated coarse-graining scheme whereby `closer' (in a suitable metric) elements are progressively merged together. This operation defines the so-called RG flow, which turns out to be deeply related to the phase diagram of the system under study \cite{wilson1983renormalization, goldenfeld2018lectures}.
Traditionally, RG techniques are based on the availability of a valid geometric embedding, i.e. a metric distance defining `how close' the elements of the system being coarse-grained onto the renormalized elements at the next hierarchical level should be, and identifying possible characteristic lengths and scaling properties in the system.
In recent years, however, researchers have identified several examples of complex systems that do not necessarily have an explicit geometric embedding: rather, these systems are naturally and more generally described by graphs where the fundamental elements are represented as vertices and their interactions as edges. 
The lack of guarantee of an explicit geometric embedding in real-world complex networks implies that vertices do not necessarily have well-defined coordinates that can be used to define the renormalization procedure in the same way as usually done for regular geometric lattices. 
Moreover, the small-world property (i.e., the fact that most pairs of nodes in real networks are connected by short paths) implies that path lengths cannot be conveniently used as metric distances, as they are a poor source of variability. 

Previous efforts to extend the notion of renormalization to complex networks have therefore explored different ideas.
The simplest approach is that of looking for (possibly hierarchically nested) communities of more densely interconnected nodes~\cite{fortunato2007resolution,fortunato2010community}, which looks at modularity (a notion of `over-density of links' with respect to a given null model) as a criterion to identify the optimal partition of nodes.
Other early attempts~\cite{gallos2007review,song2006origins} are based on a box-covering procedure, inherited from fractal analysis, relying on the identification of distance with the length of the shortest paths between nodes.
A more recent, powerful approach relies on the optimal embedding of nodes onto a `latent' or `inferred' hyperbolic metric space~\cite{garcia2018multiscale} that can then be used to guide the renormalization procedure. 
Other approaches are based on spectral coarse-graining, in which the dimension of a network is reduced while keeping the largest eigenvalues of the associated adjacency matrix (approximately) unchanged~\cite{gfeller2007spectral,cardelli2017maximal}, or via the introduction of a Laplacian operator for graphs which induces a diffusion-based renormalization scheme~\cite{pablo}. 
Provided that the considered network exhibits a non-trivial Laplacian susceptibility, the latter method is capable of successfully identifying a spatio-temporal hierarchical organization in the network. 

While all these approaches try to identify hierarchies of partitions of nodes in a given real network, they either do not address the problem of defining a (random) network model consistent with the identified partition(s), or do so in a method-dependent way, i.e. by following the same principle used to define the coarse-graining scheme itself. 
For instance, the natural generating model in the community-detection setting is the (degree-corrected) stochastic block model where the identified blocks are given an intrinsically higher internal connection probability, while the natural generating model for the hyperbolic method is the hyperbolic model itself, where the connection probabilities are determined by the geodesic distances between the (inferred) coordinates of nodes. 
However, in general one would like to have a method-free approach to the definition of a network model consistent with a given hierarchy of partitions, for instance because the notion of coexistence of multiple scales in networks may be more general than the one defining the recipe used by the coarse-graining method itself.
Indeed, an arbitrary real-world network might not necessarily possess any of the features postulated by the coarse-graining method, in which case the inferred e.g. hyperbolic, Laplacian or community structure identified as a `best fit' may actually be an artefact of the method itself. 
Moreover, many real networks are directed, a key  property that is not naturally accounted for by geometric or Laplacian approaches.

Clearly, the most general method-free definition of a network model that is compatible with a given hierarchy of partitions is one that works for \emph{any} possible such hierarchy, for instance because it can take as input multiple partitions identified by different coarse-graining methods for the same system, or because it remains `open' to future methods based on criteria yet to be defined. Following this idea, a recent approach has introduced a random graph model whose mathematical formulation is invariant with respect to the resolution level, i.e. where the functional form of the connection probability between (groups of) nodes is guaranteed to be independent of the scale (i.e. the node partition) at which the network is observed~\cite{Elenulla}. 
This Scale-Invariant Model (SIM) allows for multiscale (i.e. encompassing arbitrarily heterogeneous node aggregations) partitions of a network, does not require the presence of e.g. communities, Laplacian properties or   (geo)metric coordinates and distances, and remains self-consistent irrespective of the topological properties of the generated networks, ranging from scale-free graphs to ordinary regular lattices.
Notably, the SIM illustrates that scale-invariant connection probabilities must depend on some (latent or empirically observed) node attributes that renormalize in a specific way under coarse-graining. Optionally, the model can incorporate the dependence on dyadic features such as membership to communities and/or metric distance, but these properties are not necessary -- confirming by the way that the notion of scale-invariance in networks is more general than those induced by community structure or geometric embedding~\cite{Elenulla}.

The original formulation of the SIM is restricted to undirected graphs, while several real-world networks, including economic~\cite{gleditsch2002expanded,garlaschelli2005structure,bardoscia2021physics,ialongo2022reconstructing}, social~\cite{scott2012social,newman2002email}, biological~\cite{Garlaschelli2004PatternsOL} and material flow~\cite{martinez2022world} networks are intrinsically directed.
In general, moving from undirected to directed networks is nontrivial, because of the pervasive property of \emph{reciprocity}, which refers to the non-random tendency of pairs of nodes to form mutual connections~\cite{Garlaschelli2004PatternsOL,garlaschelli2006multispecies}.
By varying the degree of reciprocity, one can range from undirected (perfectly reciprocal) through random (areciprocal) to totally asymmetric (perfectly antireciprocal) networks~\cite{Garlaschelli2004PatternsOL,garlaschelli2006multispecies}.
Reciprocity has been found to classify directed real-world networks into consistent classes~\cite{Garlaschelli2004PatternsOL,squartini2013reciprocity}, affect the abundances of directed triads~\cite{rectriads,squartini2012triadic,squartini2013early}, determine the spectral properties of adjacency matrices~\cite{sommers1988spectrum} and significantly impact processes taking place on networks~\cite{boguna2005generalized,nowak2005evolution,zlatic2009influence,molm2010structure,gallos2012people}. 

In light of the aforementioned results, here we extend the SIM to the realm of directed networks with reciprocity, which include the previously studied undirected case~\cite{Elenulla} as an extreme one.
It should be noted that, as mentioned above, other approaches to network renormalization do not easily lend themselves to directed networks. For instance, the geometric embedding method~\cite{garcia2018multiscale} is based on necessarily symmetric distances that, \emph{per se}, cannot naturally explain the asymmetry of directed networks (indeed, directed geometric models require additional and more sophisticated ingredients~\cite{allard2023geometric,dirgeo1,dirgeo2,dirgeo3}). 
Similarly, the spectral method~\cite{pablo} requires symmetric adjacency matrices to define a Hermitian Laplacian operator.
Importantly, we find that a naive reformulation of the original~\cite{Elenulla} SIM, which merely makes the connection probabilities asymmetrical by using two features per node, fails in reproducing various topological and spectral properties of real-world networks, precisely because it does not capture the empirical patterns of reciprocity. 
On the contrary, a non-trivial extension can replicate reciprocity and the resulting properties.

The paper is organized as follows. 
In Sec.~\ref{section: DSIM} we introduce the Directed Scale-Invariant Model (DSIM).
In Sec.~\ref{section: ER}, as an pedagogical benchmark,  we present the simplest, completely homogeneous instance of the model where all nodes are assumed to be statistically equivalent. 
Then, following the approach of \cite{Elenulla}, we distinguish between a quenched scenario, where node features are treated as deterministic, and an annealed scenario, where they are interpreted as random variables.
In the quenched scenario, the SIM can be used as a model of empirical networks where node features are observable and expected to determine the topology. Then, unlike models that are not designed to remain consistent across different resolution levels, the SIM can make predictions about network properties observed at any scale. We illustrate this procedure in Sec.~\ref{Sec:Directed-quenched}, where we provide a multiscale description of the international trade network that also captures its strongly reciprocal nature.
In the annealed scenario, the model can be used to generate random networks built from the principle of scale invariance. 
We illustrate this approach in Sec.~\ref{Sec:Directed-annealed}.
Technical details on data and calculations are provided in the accompanying Supplementary Information (SI).

\section{Construction of the Directed Scale-Invariant Model}\label{section: DSIM}
In this Section we define the DSIM as a model of random directed graphs, with nontrivial reciprocity, characterized by the property that the connection probability between (blocks of) nodes has always the same functional form, irrespective of the coarse-graining (resolution level) adopted. This means that such probability depends on the chosen aggregation of nodes only through its parameters, which will obey appropriate renormalization rules. 
Importantly, the coarse-graining can be arbitrarily heterogeneous and include  \emph{multi-scale} aggregations where certain blocks may be deliberately `small' (and even coincide with the original microscopic nodes themselves) and other blocks may be very `large' (i.e., containing several microscopic nodes).
This graph model is therefore a generalization of the undirected model in \cite{Elenulla} (whose main ingredients are recalled in SI), to which it will reduce in the extreme case of complete reciprocity, i.e. when all links are reciprocated (so that the network is effectively undirected).

\subsection{Basic quantities and definitions.}
\begin{figure}
    \centering
    \includegraphics[width=0.5\textwidth]{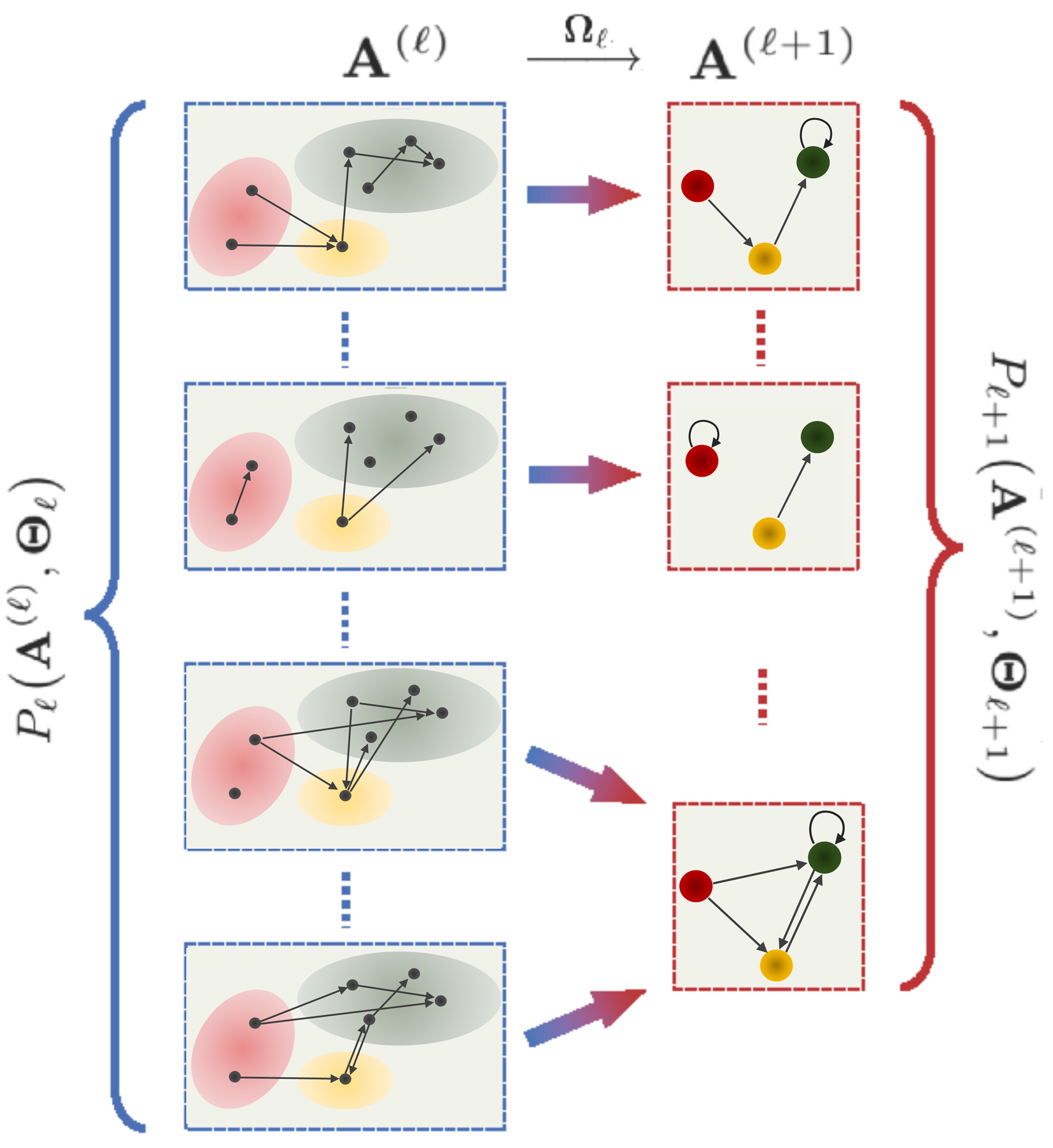}
    \caption{\small \textbf{Schematic example of the graph coarse-graining and induced ensembles in the directed case}.
    Given a probability distribution $P_\ell\big(\mathbf{A}^{(\ell)},\mathbf{\Theta}_\ell\big)$ of graphs with adjacency matrix  $\mathbf{A}^{(\ell)}$ (left), a given node partition $\mathbf{\Omega}_\ell$ is used to map sets of nodes onto `block-nodes' of the resulting coarse-grained graphs with adjacency matrix $\mathbf{A}^{(\ell+1)}$ (right).
    A directed edge from $i_{\ell+1}$ to $j_{\ell+1}$ is drawn if an edge from $i_{\ell}$ to $j_{\ell}$ is present, for any $i_{\ell} \in i_{\ell+1}, j_{\ell} \in j_{\ell+1}$.
    Note that multiple realizations of the graph at  level $\ell$ may end up in the same realization of the graph at level $\ell+1$.
    This coarse-graining will induce a new probability distribution $P_{\ell+1}\big(\mathbf{A}^{(\ell+1)},\mathbf{\Theta}_{\ell+1}\big)$.}
    \label{fig:toyCG}
\end{figure}

We consider a binary directed graph with ${N_{0}}$ `microscopic' nodes (labeled by the first $N_0$ positive integers $i_0=1,\dots, N_0$) and its ${N_{0}} \times {N_{0}}$ (Boolean and, in general, asymmetric) adjacency matrix  $\mathbf{A}^{(0)}$ with entries $a_{i_{0} j_{0}} = 1$ if a directed link from node $i_{0}$ to node $j_{0}$ exists, and $a_{i_{0} j_{0}} = 0$ otherwise. Self-loops ($a_{i_\ell i_\ell}=1$) are allowed.
For convenience, we call this microscopic graph the $0$-graph, and its $N_0$ nodes the $0$-nodes. 
First, given an \emph{arbitrary} surjective and non-overlapping partition $\Omega_0$ of the original $N_0$ $0$-nodes onto $N_1<N_0$ (coarser) nodes, labeled as $i_{1}={1},\dots,{N_1}$, we define the coarse-grained directed graph (this graph will be called the $1$-graph, and its $N_1$ nodes the $1$-nodes) with $N_1\times N_1$ adjacency matrix  $\mathbf{A}^{(1)}$ as follows.
A directed link from the $1$-node 
$i_{1}$ to the $1$-node $j_{1}$ is present if and only if, in the $0$-graph, there is at least one directed link from any $i_0\in i_1$ (i.e. from any of the $0$-nodes `inside' $i_{1}$) to any $j_{0}\in j_{1}$, as illustrated in Fig.~\ref{fig:toyCG}.
This coarse-graining step can then be iterated to identify an arbitrary sequence of coarse-grained $\ell$-graphs (each with a certain number $N_\ell$ of $\ell$-nodes), by introducing any desired hierarchy $\{\Omega_{\ell}\}_{\ell\geq 0}$ of nested (surjective and non-overlapping) partitions.
Each such $\ell$-graph will be uniquely associated to an $N_\ell\times N_\ell$ Boolean (and in general still asymmetric) adjacency matrix $\mathbf{A}^{(\ell)}$, whose entries obey the following relationship, for any $i_{\ell+1}, j_{\ell+1}$:  
\begin{equation}\label{Deq:cg rule}
    a^{(\ell +1)}_{i_{\ell +1} j_{\ell +1}} = 1 - \prod_{i_{\ell} \in i_{\ell +1}} \prod_{j_{\ell} \in j_{\ell +1}} (1- a^{(\ell )}_{i_{\ell} j_{\ell}}).
\end{equation}

Now, we consider a random graph model producing a specific realization $\mathbf{A}^{(\ell)}$ of the $\ell$-graph with probability $P_\ell\big(\mathbf{A}^{(\ell)},\mathbf{\Theta}_\ell\big)$, where $\mathbf{\Theta}_\ell$ is the set of all parameters of the model. We look for the functional form of this probability by imposing that, given a partition $\Omega_\ell$ of the $N_\ell$ $\ell$-nodes, the induced probability $P_{\ell+1}\big(\mathbf{A}^{(\ell+1)},\mathbf{\Theta}_{\ell+1}\big)$ at the next level $\ell+1$ has the same functional form as $P_\ell\big(\mathbf{A}^{(\ell)},\mathbf{\Theta}_\ell\big)$, with appropriately renormalized parameters $\mathbf{\Theta}_{\ell+1}$ (as a result, we will drop the subscript $\ell$ from the $P_\ell$ that realizes the invariance requirement).
For simplicity, we restrict ourselves to models with \emph{independent dyads} where the graph probability (for $\ell=0$, and consequently for all $\ell>0$) factorizes into dyadic connection probabilities over pairs of nodes, plus self-loop probabilities over single nodes.

Adopting the  formalism introduced in~\cite{garlaschelli2006multispecies} and reprised in~\cite{Squartini_2011,squartini2012triadic,squartini2013early,squartini2015unbiased,squartini2017maximum,gemmetto2016multiplexity}, if $\langle\cdot\rangle$ denotes an expected value with respect to the (yet to be determined) scale-invariant distribution $P\big(\mathbf{A}^{(\ell)},\mathbf{\Theta}_\ell\big)$, then the probabilities of the four possible dyads between two nodes $i_\ell$ and $j_\ell$ are: 
\begin{itemize}
\item $p_{i_\ell j_\ell}^{\rightarrow}(\mathbf{\Theta}_\ell) =  \langle a_{i_\ell j_\ell}^{\rightarrow}\rangle$ with $a_{i_\ell j_\ell}^{\rightarrow}\equiv a^{(\ell)}_{i_\ell j_\ell}(1-a^{(\ell)}_{j_\ell i_\ell}) $ for the joint probability of a single directed link from $i_\ell$ to $j_\ell$ and no reciprocal link from $j_\ell$ to $i_\ell$; 
\item $p_{i_\ell j_\ell}^{\leftarrow}(\mathbf{\Theta}_\ell) =  \langle a_{i_\ell j_\ell}^{\leftarrow}\rangle$ with $a_{i_\ell j_\ell}^{\leftarrow}\equiv a^{(\ell)}_{j_\ell i_\ell}(1-a^{(\ell)}_{i_\ell j_\ell})$ for the joint probability of a single directed link from $j_\ell$ to $i_\ell$ and no reciprocal link from $i_\ell$ to $j_\ell$; 
\item $p_{i_\ell j_\ell}^{\leftrightarrow}(\mathbf{\Theta}_\ell) =  \langle a_{i_\ell j_\ell}^{\leftrightarrow}\rangle$ with $a_{i_\ell j_\ell}^{\leftrightarrow}\equiv a^{(\ell)}_{i_\ell j_\ell}a^{(\ell)}_{j_\ell i_\ell}$ for the joint probability of two reciprocal links between $i_\ell$ and $j_\ell$; 
\item $p_{i_\ell j_\ell}^{\nleftrightarrow}(\mathbf{\Theta}_\ell) =    \langle a_{i_\ell j_\ell}^{\nleftrightarrow}\rangle$ with  $a_{i_\ell j_\ell}^{\nleftrightarrow}\equiv(1-a^{(\ell)}_{i_\ell j_\ell})(1-a^{(\ell)}_{j_\ell i_\ell}) $ for the joint probability of no links between $i_\ell$ and $j_\ell$.
\end{itemize}
Clearly, 
$p_{i_\ell j_\ell}^{\rightarrow}(\mathbf{\Theta}_\ell) + p_{i_\ell j_\ell}^{\leftarrow}(\mathbf{\Theta}_\ell) + p_{i_\ell j_\ell}^{\leftrightarrow}(\mathbf{\Theta}_\ell) + p_{i_\ell j_\ell}^{\nleftrightarrow}(\mathbf{\Theta}_\ell) = 1$.
In order to generate reciprocity, i.e. the nontrivial occurrence of mutual links between the same two nodes as observed in virtually all real-world directed networks~\cite{Garlaschelli2004PatternsOL,squartini2013reciprocity}, we will not assume that the joint probabilities further factorize over the marginal probabilities $p_{i_\ell j_\ell}(\mathbf{\Theta}_\ell)\equiv\langle a^{(\ell)}_{i_\ell j_\ell}\rangle=p^\rightarrow_{i_\ell j_\ell}(\mathbf{\Theta}_\ell)+p^\leftrightarrow_{i_\ell j_\ell}(\mathbf{\Theta}_\ell)$ for individual edges.
For self-loops, the only relevant probability is $p_{i_\ell i_\ell}(\mathbf{\Theta}_\ell)\equiv\langle a^{(\ell)}_{i_\ell i_\ell}\rangle=p^\leftrightarrow_{i_\ell i_\ell}(\mathbf{\Theta}_\ell)=1-p^\nleftrightarrow_{i_\ell i_\ell}(\mathbf{\Theta}_\ell)$ since $p^\rightarrow_{i_\ell i_\ell}(\mathbf{\Theta}_\ell)=p^\leftarrow_{i_\ell i_\ell}(\mathbf{\Theta}_\ell)=0$ (self-loops necessarily reciprocate).

Before looking for the scale-invariant functional form of the above probabilities, we assume that the latter may depend on \emph{global} (scalar) parameters $\epsilon$, $\delta$, $\eta$ (determining, as we will see, the overall numbers of directed links, reciprocated links and self-loops respectively), \emph{node-specific} ($N_\ell$-dimensional) features $\{x_{i_\ell}\}_{i_\ell=1}^{N_\ell}$, $\{y_{i_\ell}\}_{i_\ell=1}^{N_\ell}$, $\{z_{i_\ell}\}_{i_\ell=1}^{N_\ell}$, $\{w_{i_\ell}\}_{i_\ell=1}^{N_\ell}$ called \emph{fitness} values (separately determining the intrinsic tendency of individual nodes of forming reciprocated pairs of links, unreciprocated out-going links, unreciprocated in-coming links and self-loops), and finally \emph{dyadic} factors parametrized by an $N_\ell\times N_\ell$ matrix $\mathbf{D}^{(\ell)}$ (determining pairwise connection preferences not attributable to global or node-specific ones, e.g. similarities, distances, membership to communities, etc.). While we may in principle add higher-order parameters (controlling for triadic preferences and so on), this would not be natural for a model conceived with independent dyads, as already noted in~\cite{Elenulla}.
Having clarified our parametrization, and to ease the notation, from now on we omit the dependence on $\mathbf{\Theta}_\ell$ in all quantities. 

\subsection{Scale-invariant connection probabilities.}
We can now look for the scale-invariant form of all probabilities. First, we demand that the undirected projection of our directed model reduces to the scale-invariant model considered in~\cite{Elenulla} for undirected graphs (without this constraint, the directed model we are building here would violate scale-invariance when projected onto its undirected representation). We can therefore  constrain the functional form of the `undirected' connection probability
\begin{eqnarray}\label{eq:defq}
    q_{i_\ell j_\ell}\equiv p^\rightarrow_{i_\ell j_\ell}+p^\leftarrow_{i_\ell j_\ell}+p^\leftrightarrow_{i_\ell j_\ell}=1-p^\nleftrightarrow_{i_\ell j_\ell}
\end{eqnarray} 
(i.e. the marginal probability of nodes $i_\ell$ and $j_\ell$ being connected by at least one link, whatever the direction) to have the form already derived in~\cite{Elenulla}, i.e.:
\begin{equation}
  q_{i_\ell j_\ell}=
  \begin{cases}
               1 - e^{-\delta x_{i_\ell} x_{j_\ell}  f(d_{i_\ell j_\ell})} & i_\ell\neq j_\ell\\
              1 - e^{-\frac{\delta}{2} x_{i_\ell}^2 f(d_{i_\ell i_\ell}) - \eta w_{i_\ell} } & i_\ell = j_\ell
        \end{cases}
\label{q}
\end{equation}
where $\delta$, $\{x_{i_\ell}\}_{i_\ell=1}^{N_\ell}$, $\{w_{i_\ell}\}_{i_\ell=1}^{N_\ell}$, $\{d_{i_\ell j_\ell}\}_{i_\ell,j_\ell=1}^{N_\ell}$ are all positive and $f$ is a positive monotonic function. The global parameter $\delta$ tunes the overall density of \emph{undirected} links, the node-specific fitness $x_{i_\ell}$ controls the tendency of node $i_\ell$ of forming undirected (i.e. of any directionality) links, and $f$ incorporates the (optional) effects of the dyadic properties $\{d_{i_\ell j_\ell}\}_{i_\ell,j_\ell=1}^{N_\ell}$. As an addition to the undirected model in~\cite{Elenulla}, we have introduced an extra fitness $w_{i_\ell}$ (coupled to another global parameter $\eta$) separately controlling for the tendency of $i_\ell$ of forming a self-loop more (if $\eta>0$) or less (if $\eta<0$) likely than otherwise determined  by $x_{i_\ell}$ (note that \emph{self-loops are intrinsically reciprocated}).
To guarantee a positive $q_{i_\ell i_\ell}$, one must require 
\begin{equation}
\eta\ge -\min_{i_\ell}\left\{ \frac{\delta x_{i_\ell}^2 f(d_{i_\ell i_\ell})}{2 w_{i_\ell}}\right\}.
\label{cond1}
\end{equation}
Also note that, in order to preserve the consistency between the directed network and its undirected projection, we need the matrix $\mathbf{D}^{(\ell)}$ to be symmetric ($d_{i_\ell j_\ell} = d_{j_\ell i_\ell}$) as in the undirected scenario~\cite{Elenulla}. The key property of the probability $q_{i_\ell j_\ell}$ in~\eqref{q} is that it preserves its functional form under arbitrary aggregations: if the coarse-grained network at the next level $\ell+1$ is considered, then the probability $q_{i_{\ell+1} j_{\ell+1}}$ that the two $(\ell+1)$-nodes $i_{\ell+1}$ and $j_{\ell+1}$ are connected by a link (i.e. that any pair of their constituent $\ell$-nodes are connected) is given by exactly the same expression, with parameters $\delta$, $\eta$ remaining unchanged and the other ones renormalizing as~\cite{Elenulla}
\begin{eqnarray}
&x_{i_{{\ell}+1}} \equiv \sum_{i_{{\ell}} \in i_{{\ell}+1}}  x_{i_{{\ell}}},\quad w_{i_{{\ell}+1}} \equiv \sum_{i_{{\ell}} \in i_{{\ell}+1}}  w_{i_{{\ell}}},&\label{renormxw}\\
&f(d_{i_{{\ell}+1} j_{{\ell}+1}})\equiv \frac{\sum_{i_{{\ell}} \in i_{{\ell}+1}}\sum_{j_{{\ell}}\in j_{{\ell}+1}} x_{i_{{\ell}}} x_{j_{{\ell}}} f(d_{i_{{\ell}} j_{{\ell}}})}{\sum_{i_{{\ell}} \in i_{{\ell}+1}}x_{i_{{\ell}}}~\sum_{j_{{\ell}} \in j_{{\ell}+1}}x_{j_{{\ell}}}}.&\label{renormfxx}
\end{eqnarray}
Note that~\eqref{renormfxx} holds also for self-distances ($i_{{\ell}+1}=j_{{\ell}+1}$).

Next, we consider the other marginal, but `directed', probability $p_{i_\ell j_\ell}$ introduced above, representing the probability of a directed edge being present from $i_\ell$ to $j_\ell$, irrespective of the presence of the edge in the opposite direction. By demanding that $p_{i_\ell j_\ell}$ fulfils the same scale-invariance requirement that in the undirected case leads to~\eqref{q}, we arrive at the expression
\begin{equation} \label{Deq: p_ij ignorante}
    p_{i_\ell j_\ell}=\begin{cases}
                1 - e^{-\epsilon y_{i_\ell} z_{j_\ell} f(d_{i_\ell j_\ell})}   & i_\ell\neq j_\ell\\
              1 - e^{-\frac{\delta}{2} x_{i_\ell}^2 f(d_{i_\ell i_\ell}) - \eta w_{i_\ell} } & i_\ell = j_\ell
        \end{cases}
\end{equation}
where $\epsilon>0$ tunes the overall density of directed links and, because of the possible asymmetry of $p_{i_\ell j_\ell}$ for $i_\ell\ne j_\ell$, we have introduced
\emph{two} sets of fitness variables $\{y_{i_{\ell}}\}_{i_{\ell} = 1}^{N_{\ell}}$ and $\{z_{i_{\ell}}\}_{i_{\ell} = 1}^{N_{\ell}}$ (representing the intrinsic tendency of each node of establishing out-going and in-coming links respectively), while for $i_\ell= j_\ell$ we have necessarily enforced $p_{i_\ell i_\ell}\equiv q_{i_\ell i_\ell}$. 
Upon coarse-graining, $\epsilon$ is unchanged while, in analogy with~\eqref{renormxw} and~\eqref{renormfxx}, the other parameters renormalize as 
\begin{eqnarray}
&y_{i_{{\ell}+1}} \equiv \sum_{i_{{\ell}} \in i_{{\ell}+1}}  y_{i_{{\ell}}},\quad z_{i_{{\ell}+1}} \equiv \sum_{i_{{\ell}} \in i_{{\ell}+1}}  z_{i_{{\ell}}},&\label{renormyz}\\
&f\big(d_{i_{{\ell}+1} j_{{\ell}+1}}\big)\equiv \frac{\sum_{i_{{\ell}} \in i_{{\ell}+1}}\sum_{j_{{\ell}}\in j_{{\ell}+1}} y_{i_{{\ell}}} z_{j_{{\ell}}} f\big(d_{i_{{\ell}} j_{{\ell}}}\big)}{\sum_{i_{{\ell}} \in i_{{\ell}+1}}y_{i_{{\ell}}}~\sum_{j_{{\ell}} \in j_{{\ell}+1}}z_{j_{{\ell}}}}.&\label{renormfyz}
\end{eqnarray}
Note that~\eqref{renormfyz} is not to be applied when $i_{\ell+1}= j_{\ell+1}$, in which case the requirement of  scale-invariance bounces back to the undirected probabilities in~\eqref{eq:defq} and thus to the renormalization rules in~\eqref{renormxw} and~\eqref{renormfxx}. 
On the contrary, when $i_{\ell+1} \neq j_{\ell+1}$, Eqs~(\ref{renormfxx})-(\ref{renormfyz}) should be realized simultaneously. 
To ensure this, a convenient sufficient condition is that the dyadic factors $\{d_{i_\ell j_\ell}\}_{i_\ell,j_\ell=1}^{N_\ell}$ are \emph{ultrametric} (or more precisely \emph{metaultrametric}) distances compatible with the chosen hierarchy $\{\Omega_\ell\}_{\ell\ge 0}$ of nested partitions. 
\emph{Metric} distances satisfy four axioms, for all $i,j$: positivity ($d_{ij}> 0$ if $i\ne j$), symmetry ($d_{ij}=d_{ji}$), triangular inequality ($d_{ij} \leq d_{ik}+d_{jk}$ $\forall k$) and null self-distance ($d_{ij}=0$ iff $i=j$).
\emph{Metametric}~\cite{metametric,das2017geometry} (also called \emph{partial metric}~\cite{bukatin2009partial} or \emph{dislocated}~\cite{hitzler2000dislocated}) distances relax the last axiom by requiring only $i=j$ if $d_{ij}=0$, so they admit positive self-distance ($d_{ii}\ge 0$).
\emph{Ultrametric} distances~\cite{rammal1986ultrametricity} are metric distances that satisfy a stronger version of the triangular inequality, known as ultrametric inequality ($d_{ij} \leq \max_{k} \{d_{ik},d_{jk}\}$ $\forall k$). As a result, they can be arranged on a dendrogram whereby the distance between two `leaves' equals the height of the closest branching point between them.
In this paper, we call \emph{metaultrametric} distance a metametric that satisfies the ultrametric inequality. 
If $d$ is metaultrametric over the dendrogram induced by the nested hierarchy $\{\Omega_\ell\}_{\ell\ge 0}$, then $f(d_{i_\ell j_\ell})$ comes out of the sums in~\eqref{renormfxx} and~\eqref{renormfyz} and is therefore invariant, for any assignment of the fitness:
\begin{equation} \label{eq:SIf}
f(d_{i_{{\ell +1}} j_{{\ell}+1}}) =f(d_{i_{{\ell}} j_{{\ell}}}) \quad \textrm{if}\; \;i_{{\ell}}\in i_{{\ell}+1},\; j_{{\ell}}\in j_{{\ell}+1},
\end{equation}
with $i_{{\ell}+1} \neq j_{{\ell}+1}$. For $i_{{\ell}+1} = j_{{\ell}+1}$, $f(d_{i_{{\ell +1}} j_{{\ell}+1}})$ is still obtained through~\eqref{renormfxx}.
Allowing $d_{i_\ell i_\ell}>0$ enables non-zero self-interactions ($p_{i_\ell i_\ell}>0$) even when $\eta=0$. This is particularly relevant for coarse-grained configurations, because any block of nodes connected by at least one link has a self-loop.  From now on, we assume that the distances $\{\mathbf{D}^{(\ell)}\}_{\ell\ge 0}$ in our model are metaultrametric over $\{\Omega_\ell\}_{\ell\ge 0}$.
Note that this requirement is unnecessary, although convenient, in the undirected case~\cite{Elenulla}.

By rearranging the marginal probabilities $q_{i_\ell j_\ell}$ and $p_{i_\ell j_\ell}$ in~\eqref{q} and~\eqref{Deq: p_ij ignorante} respectively, we finally arrive at the fundamental expressions for the joint probabilities:
\begin{eqnarray}
\small
        p_{i_\ell j_\ell }^{\rightarrow}
        \!\!&=&\!\!\begin{cases}
            e^{-\epsilon z_{i_\ell} y_{j_\ell} f(d_{i_\ell j_\ell})}- e^{-\delta x_{i_\ell } x_{j_\ell } f(d_{i_\ell j_\ell})} &i_\ell  \neq j_\ell \\
            0 &i_\ell  = j_\ell 
        \end{cases},\label{pright} \\
        p_{i_\ell j_\ell }^{\leftarrow} \!\!&=&\!\! \begin{cases}
             e^{-\epsilon y_{i_\ell } z_{j_\ell } f(d_{i_\ell j_\ell }) }- e^{-\delta x_{i_\ell } x_{j_\ell } f(d_{i_\ell j_\ell })} &i_\ell  \neq j_\ell \\
            0 &i_\ell  = j_\ell 
        \end{cases},\label{pleft}\\
        p_{i_\ell j_\ell }^{\nleftrightarrow} \!\!&=&\!\! \begin{cases}
              e^{-\delta x_{i_\ell} x_{j_\ell } f(d_{i_\ell j_\ell })} & i_\ell  \neq j_\ell \\
               e^{-\frac{\delta}{2} x_{i_\ell }^2 f(d_{i_\ell i_\ell }) - \eta w_{i_\ell } } & i_\ell  = j_\ell 
        \end{cases},\label{pno} \\
        p_{i_\ell j_\ell }^{\leftrightarrow}   \!\!&=&\!\! \begin{cases}
              1 - p_{i_\ell j_\ell }^{\rightarrow}-p_{i_\ell j_\ell }^{\leftarrow}-p_{i_\ell j_\ell }^{\nleftrightarrow} & i_\ell \neq j_\ell \\
              1 - e^{-\frac{\delta}{2} x_{i_\ell }^2 f(d_{i_\ell i_\ell }) - \eta w_{i_\ell } } & i_\ell  = j_\ell 
        \label{pboth}\end{cases}.
\end{eqnarray}
Note that the latter expression implies, whenever $i_\ell \neq j_\ell$,
$p_{i_\ell j_\ell }^{\leftrightarrow}  = 1 - e^{-\epsilon y_{i_\ell } z_{j_\ell } f(d_{i_\ell j_\ell }) }- e^{-\epsilon z_{i_\ell } y_{j_\ell } f(d_{i_\ell j_\ell })} + e^{-\delta x_{i_\ell } x_{j_\ell } f(d_{i_\ell j_\ell })}$.
Also note that, in analogy with~\eqref{cond1}, the following condition is needed to restrict all probabilities within the interval $[0,1]$:
\begin{equation} \label{eq: CE}
    \delta^{(i_\ell,j_\ell)}_\textrm{min} \leq \delta  \leq \delta^{(i_\ell,j_\ell)}_\textrm{max}   \quad \forall i_\ell \neq j_\ell
\end{equation}
where $\delta^{(i_\ell,j_\ell)}_\textrm{max} \equiv - \frac{\ln{ \left( e^{-\epsilon y_{i_\ell} z_{j_\ell} f(d_{i_\ell j_\ell})} + e^{-\epsilon y_{j_\ell} z_{i_\ell} f(d_{i_\ell j_\ell})} - 1 \right)}}{x_{i_\ell} x_{j_\ell} f(d_{i_\ell j_\ell}) } $ if $ (1-p_{i_\ell j_\ell}-p_{j_\ell i_\ell})  > 0$ and $\delta^{(i_\ell,j_\ell)}_\textrm{max} \equiv +\infty$ otherwise, while $\delta^{(i_\ell,j_\ell)}_\textrm{min} \equiv\frac{ \epsilon \max{(y_{i_\ell} z_{j_\ell}, y_{j_\ell} z_{i_\ell})}}{x_{i_\ell} x_{j_\ell}}$. Note that the above condition should hold for all pairs $i_\ell$, $j_\ell$ simultaneously, which also means for all hierarchical levels $\ell\ge 0$. Whether it is possible to fulfil the condition therefore depends not only on the values of $\{x_{i_\ell}\}_{i_\ell=1}^{N_\ell}$, $\{y_{i_\ell}\}_{i_\ell=1}^{N_\ell}$, $\{z_{i_\ell}\}_{i_\ell=1}^{N_\ell}$, $\{w_{i_\ell}\}_{i_\ell=1}^{N_\ell}$ of all nodes for a given hierarchical level $\ell$, but also on the entire hierarchy $\{\Omega_{\ell}\}_{\ell\geq 0}$ of chosen partitions. Sufficient, although not necessary, conditions are given in SI.

We highlight two important quantities characterizing the model. The first, local property is the \emph{conditional} probability 
\begin{equation} \label{eq:conditional}
r_{i_\ell j_\ell}\equiv \textrm{Prob}(i_\ell \rightarrow j_\ell | j_\ell \rightarrow i_\ell)= \frac{ p_{i_\ell j_\ell}^{\leftrightarrow}}{p_{j_\ell i_\ell}}
\end{equation} 
that a link from $i_\ell$ to $j_\ell$ exists, given that the reciprocal link from $j_\ell$ to $i_\ell$ exists. Note that $r_{i_\ell i_\ell}=1$, i.e. a self-loop is necessarily reciprocated (by itself).
The second, global property is the overall \emph{reciprocity} $\langle r \rangle$ defined as the ratio of the expected value of the number 
$L^{\leftrightarrow} \equiv \sum_{i_\ell=1}^{N_\ell}\sum_{j_\ell\neq i_\ell} a^{(\ell)}_{i_\ell j_\ell}a^{(\ell)}_{j_\ell i_\ell}$ of reciprocated links to the expected value of the number
$L \equiv \sum_{i_\ell=1}^{N_\ell}\sum_{j_\ell\neq i_\ell}a^{(\ell)}_{i_\ell j_\ell}$ of links in total, i.e.
\begin{equation}\label{eq: reciprocity}
    \langle r \rangle \equiv \frac{\langle L^{\leftrightarrow} \rangle }{\langle L \rangle } =\frac{\sum_{i_\ell=1}^{N_\ell}\sum_{j_\ell\neq i_\ell}p_{i_\ell j_\ell}^{\leftrightarrow}}{\sum_{i_\ell=1}^{N_\ell}\sum_{j_\ell\neq i_\ell}p_{i_\ell j_\ell}}.
\end{equation}
In the special case where all nodes are statistically equivalent (i.e., they are assigned the same fitness values), all probabilities do not depend on the specific pair of nodes and in particular $r_{i_\ell j_\ell}\equiv r=\langle r \rangle$ which equals the global reciprocity~\cite{Garlaschelli2004PatternsOL}, motivating the choice of a common symbol for the two quantities.

\subsection{Relevant cases.} \label{parag:relevant}
The above model generalizes the undirected model considered in~\cite{Elenulla} to the directed case. Before discussing its use in more general settings, we briefly discuss some relevant cases that represent useful benchmarks for later.  

\begin{itemize}
\item \emph{Maximal reciprocity} ($r_{i_\ell j_\ell}=1$) is achieved when all the connections are bidirectional by construction and reduces to the undirected case already discussed in~\cite{Elenulla}, with the additional freedom on controlling self-loops separately from the other links. In particular, we get $q_{i_\ell j_\ell} =  p_{i_\ell j_\ell} = p_{j_\ell i_\ell}=p_{i_\ell j_\ell}^{\leftrightarrow}$ and $p_{i_\ell j_\ell}^{\rightarrow} = p_{j_\ell i_\ell}^{\leftarrow} = 0 \; \forall \, i_\ell,j_\ell$.
Clearly, in this case the conditions in~\eqref{eq: CE} force the fitness variables to be related via the expression $y_{i_\ell } \equiv z_{i_\ell } \equiv x_{i_\ell }\sqrt{{\delta}/{\epsilon}}  \; \forall \, i_\ell $.

\item \emph{Positive reciprocity} ($r_{i_\ell j_\ell}> p_{i_\ell j_\ell}$) is achieved when the network has a preference for the creation of reciprocal links, even when not maximally reciprocated.
In particular, the conditional reciprocation probability $r_{i_\ell j_\ell}$ is larger than the unconditional connection probability $p_{i_\ell j_\ell}$. Equivalently, the joint probability of a pair of links being reciprocated is larger than the probability of the same event occurring by chance if the two links were independent: $p_{i_\ell, j_\ell}^{\leftrightarrow} > p_{i_\ell j_\ell} p_{j_\ell i_\ell}$. Refs.~\cite{Garlaschelli2004PatternsOL,garlaschelli2006multispecies} report several real-world networks (in decreasing order of reciprocity, the World Trade Web, the World Wide Web, neural networks, email networks, 
word networks and metabolic networks) as belonging to this class of positive reciprocity.

\item \emph{Random reciprocity or areciprocity} ($r_{i_\ell j_\ell}= p_{i_\ell j_\ell}$) is achieved in the `neutral' or trivial case when the two links between any two nodes are independent. Reciprocity is therefore the result of sheer chance, given the marginal connections probabilities: $p_{i_\ell, j_\ell}^{\leftrightarrow} = p_{i_\ell j_\ell} p_{j_\ell i_\ell}$.
This case is achieved by the condition $\delta x_{i_\ell} x_{j_\ell} = \epsilon (y_{i_\ell} z_{j_\ell} + y_{j_\ell} z_{i_\ell})  \;  \forall \, i_\ell, j_\ell$.
This condition can be seen as an equivalent way to discriminate between a {reciprocal} ensemble where $\delta x_i x_j < \epsilon (y_i z_j + y_j z_i)$, and an {antireciprocal} ensemble (see below) where $\delta x_i x_j > \epsilon (y_i z_j + y_j z_i)$. The areciprocal case is useful to define the reference `random' reciprocity 
\begin{equation} \label{eq: rec_rand}
    \langle r \rangle_\textrm{rand} = \frac{\sum_{i_\ell=1}^{N_\ell}\sum_{j_\ell\neq i_\ell}p_{i_\ell j_\ell}p_{j_\ell i_\ell}}{\sum_{i_\ell=1}^{N_\ell}\sum_{j_\ell\neq i_\ell}p_{i_\ell j_\ell}}
\end{equation} 
as the value achieved by~\eqref{eq: reciprocity} in a network where no dependency between mutual links is present.

\item \emph{Negative reciprocity or antireciprocity} ($r_{i_\ell j_\ell}< p_{i_\ell j_\ell}$) is achieved when the network tries to avoid the creation of reciprocal links.
The conditional reciprocation probability $r_{i_\ell j_\ell}$ is in this case smaller than the unconditional connection probability $p_{i_\ell j_\ell}$. Equivalently, the joint probability of a pair of links being reciprocated is smaller than the probability of the same event occurring by chance if the two links were independent: $p_{i_\ell, j_\ell}^{\leftrightarrow} < p_{i_\ell j_\ell} p_{j_\ell i_\ell}$.
Ref.~\cite{Garlaschelli2004PatternsOL} reports several real-world networks (e.g. corporate shareholding networks and food webs) as belonging to this class of negative reciprocity.

\item \emph{Minimal reciprocity (maximal antireciprocity)} corresponds, ideally, to a complete absence of mutual links, where all  connections are either missing or unidirectional so to have $r_{i_\ell j_\ell}= p_{i_\ell j_\ell}^{\leftrightarrow} = 0 $, $p_{i_\ell j_\ell}^{\rightarrow} = p_{i_\ell j_\ell}$, $p_{j_\ell i_\ell}^{\leftarrow} = p_{j_\ell i_\ell}  \; \forall \, i_\ell, j_\ell$. However, this extreme case cannot always be reached, as there might be no unique choice for $\delta$ achieving the fully antireciprocal limit $\delta =\delta_{i_\ell j_\ell}^\textrm{max}$ for all node pairs simultaneously. This becomes increasingly difficult as the distribution of the fitness variables becomes broader (and, in general, as more arbitrary coarse-grainings are iterated): if $1-p_{i_\ell j_\ell}-p_{j_\ell i_\ell} < 0$, then $p_{i_\ell j_\ell}^{\leftrightarrow}>0$ irrespective of $\delta$, since $p_{i_\ell j_\ell}^{\leftrightarrow} = p_{i_\ell j_\ell}+p_{j_\ell i_\ell}-q_{i_\ell j_\ell} > 1 - q_{i_\ell j_\ell} \geq 0$.
This effect will be discussed more transparently in Sec.~\ref{section: ER}.

\end{itemize}

\subsection{Graph probability and scale invariance.} \label{Dparag: Renormalization of the DSIM}

Having derived all the dyadic connection probabilities in~\eqref{pright}~-~\eqref{pboth}, we can now use them to obtain the full probability $P(\mathbf{A}^{(\ell)})$ for the entire graph $\mathbf{A}^{(\ell)}$ at any hierarchical level $\ell$. To do so, it is convenient to represent the graph through the mutually exclusive dyads $a_{i_\ell j_\ell}^{\rightarrow}$, $a_{i_\ell j_\ell}^{\leftarrow}$, $a_{i_\ell j_\ell}^{\leftrightarrow}$, $a_{i_\ell j_\ell}^{\nleftrightarrow}$ introduced above: 
\begin{eqnarray}\label{Deq_product2}
\small
P(\mathbf{A}^{(\ell)}) \! &=& \!\! \prod_{i_{\ell}=1}^{N_{\ell}}  \prod_{j_{\ell}=1}^{i_{\ell}} (p_{i_{\ell}j_{\ell}}^{\rightarrow} )^{{a_{i_{\ell}j_{\ell}}^{\rightarrow}} } (p_{i_{\ell}j_{\ell}}^{\leftarrow})^{a_{i_{\ell}j_{\ell}}^{\leftarrow}}(p_{i_{\ell}j_{\ell}}^{\leftrightarrow})^{a_{i_{\ell}j_{\ell}}^{\leftrightarrow}}(p_{i_{\ell}j_{\ell}}^{\nleftrightarrow})^ {a_{i_{\ell}j_{\ell}}^{\nleftrightarrow} }\nonumber\\
&\equiv&\!
  \frac{e^{-\mathcal{H}^{(\ell)}_\textrm{eff}\left(\mathbf{A}^{(\ell)} \right)}}{\mathcal{Z}^{(\ell)}},
\end{eqnarray}
where we have introduced the \emph{effective Hamiltonian}
\begin{eqnarray}\label{eq:Heff0}
\mathcal{H}^{(\ell)}_\textrm{eff}(\mathbf{A}^{(\ell)} ) = &-&\!\! \sum_{i_\ell =1}^{N_\ell}    \sum_{j_\ell =1}^{i_\ell}  \Big[ a^{\rightarrow}_{i_\ell j_\ell} \ln{  \frac{p^{\rightarrow}_{i_\ell j_\ell}}{p^{\nleftrightarrow}_{i_\ell j_\ell}}} \\
&+&\!\!   a^{\leftarrow}_{i_\ell j_\ell} \ln{\frac{p^{\leftarrow}_{i_\ell j_\ell}}{p^{\nleftrightarrow}_{i_\ell j_\ell}}} \,+\, a^{\leftrightarrow}_{i_\ell j_\ell} \ln{ \frac{p^{\leftrightarrow}_{i_\ell j_\ell}}{p^{\nleftrightarrow}_{i_\ell j_\ell}}} \Big]\nonumber
\end{eqnarray}
and the \emph{partition function}
\begin{eqnarray}
\mathcal{Z}^{(\ell)}  \!&\equiv&\!\!\sum_{\{\mathbf{A}^{(\ell)}\}} e^{-\mathcal{H}^{(\ell)}_\textrm{eff}\left(\mathbf{A}^{(\ell)} \right)}\,=\,\prod_{i_\ell =1}^{N_\ell}    \prod_{j_\ell =1}^{i_\ell}   \frac{1 }{p^{\nleftrightarrow}_{i_\ell j_\ell}}  \\
    \!\!&=& 1-p_{i_{\infty}i_{\infty}}\,=\, e^{\frac{\delta}{2}  x_{i_{\infty}}^2 f(d_{i_{\infty},i_{\infty}}) +\eta  w_{i_{\infty}} } ,
\end{eqnarray}

with $x_{i_\infty}\equiv\sum_{i_\ell=1}^{N_\ell}x_{i_\ell}$, $w_{i_{\infty}} \equiv\sum_{i_{\ell}=1}^{N_{\ell}} w_{i_{\ell}}$ and $f(d_{i_\infty,i_\infty})\equiv{x^{-2}_{i_\infty}}\sum_{i_{\ell}=1}^{N_\ell}\sum_{j_{\ell}=1}^{N_\ell} {x_{i_{\ell}} x_{j_{\ell}} f(d_{i_{\ell},j_{\ell}})}$ being scale-invariant quantities: $f(d_{i_\infty,i_\infty})$ is independent of the hierarchical level $\ell$, while $x_{i_\infty}$ and $w_{i_\infty}$ are even independent of the partitions (see SI).
As a result, the partition function is exactly scale-invariant itself, as in Kadanoff's real-space renormalization~\cite{kadanoff2000statistical}.

\section{A simplified benchmark: the homogeneous case} \label{section: ER}
As the simplest benchmark for the more complicated cases that we will consider later, we first illustrate the purely homogeneous case where the fitness variables are all equal and the dyadic factors are switched off ($f \equiv 1$), so that the only free parameters left are $\epsilon$ and $\delta$. Without loss of generality, we can then set $x_{i_0} \equiv y_{i_0} \equiv z_{i_0} \equiv w_{i_0} = 1\;  \forall \, {i_0}=1,N_0$ at level $\ell=0$. Homogeneity will then be preserved at any subsequent level $\ell>0$, provided that blocks of nodes are all equal in size for any $\ell$. This model represents a scale-invariant reparametrization of the well known \emph{p1 model} by Holland and Leinhard~\cite{holland1981exponential}, where the random graph is homogeneous but with nontrivial reciprocity. 
We do not consider self-loops at level $\ell=0$ and we therefore set $\eta = -\delta/2$, so that $p^\leftrightarrow_{{i_0} {i_0}}=p^\rightarrow_{{i_0}{i_0}}=p^\leftarrow_{{i_0}{i_0}}=p_{{i_0}{i_0}}=q_{{i_0}{i_0}}=0$ for all ${i_0}=1,N_0$.
In this scenario, all the connection probabilities are identical for different pairs of nodes $({i_0},{j_0})$ (with ${i_0} \neq {j_0})$ and equal to
 \begin{equation} \label{eq: all probs in ER}
 \begin{split}
    &p = 1 - e^{-\epsilon},\\
    &q = 1 - e^{-\delta},\\
    &p^{\rightarrow}   = q - p = e^{-\epsilon} - e^{-\delta},\\
    &p^{\leftrightarrow}= 2p -q = 1 +  e^{-\delta} - 2e^{-\epsilon}  .
 \end{split}
 \end{equation}
Notably, the conditions in~\eqref{eq: CE} become very transparent:
 \begin{equation*}
     \epsilon\leq \delta \leq \delta_{\max}, \quad \delta_{\max} = \begin{cases}
         - \ln{\left(2 e^{-\epsilon} -1\right)}, &\text{if }  p < \frac{1}{2}\\
        \infty, &\text{otherwise}
     \end{cases}
 \end{equation*}
and are \emph{global} (i.e. not pair-specific). This allows us to easily interpret the two extreme values of $\delta$ in terms of the resulting expected reciprocity, which in this case equals $\langle r \rangle={p^{\leftrightarrow}}/{p}$.
In particular, for $\delta = \delta_{\min}=\epsilon$ we recover maximal reciprocity ($\langle r \rangle = 1$), while for $\delta = \delta_{\max}$ we recover minimal reciprocity ($\langle r \rangle$= $\langle r \rangle_{\min}$).
In the latter case we have to distinguish between two possibilities: if $p > \frac{1}{2}$, then $\langle r \rangle_{\min}$   is necessarily larger than zero since $p^{\leftrightarrow} ( \delta_{\max}) >0$. 
This is correctly mirroring the fact that if the average link density (which in the homogeneous case equals $p$) is larger than $1/2$, then the minimum number $L^{\leftrightarrow}_{\min}$ of reciprocated links is given by twice the number of links exceeding the number of vertex pairs, i.e. $L^{\leftrightarrow}_{\min} = 2\left[ L - {N(N-1)}/{2}\right]$. 
This implies that the realized value of $r$ cannot be smaller than ${L^{\leftrightarrow}_{\min}}/{L}$ and, consequently, the minimum expected reciprocity takes the strictly positive value $\langle r \rangle_{\delta_{\max}}  = 2 - \frac{1}{p} $.  
On the other hand, for $p \leq 1/2$ we can reach the complete absence of reciprocal links and get $\langle r \rangle_{\delta_{\max}}  = 0$ for  $\delta_{\max} = - \log{(1-2p)}$.

In what follows, we keep discussing the properties of the model at level $\ell=0$. Clearly, the properties at the next levels $\ell>0$ will depend on the details of the partitions chosen (in any case, self-loops become more likely in coarser configurations). Importantly, choosing block nodes of the same size at each level will keep the model homogeneous, with the same values of the global parameters $\delta$, $\epsilon$ and $\eta$, but rescaled and increasingly larger values of the fitnesses (yet still equal for different blocks). In the latter case, adjusting the results of this section to any desired hierarchical level is straightforward: coarser (finer) representations of the network effectively correspond to larger (smaller) values of the global parameters. Equivalently, studying the model for different values of the parameters can be reinterpreted as considering different levels of coarse-grainings. By contrast, partitions into heterogeneously sized blocks will effectively turn the model into a non-homogeneous one.

\begin{figure}[t]
    \includegraphics[width=0.39\textwidth]{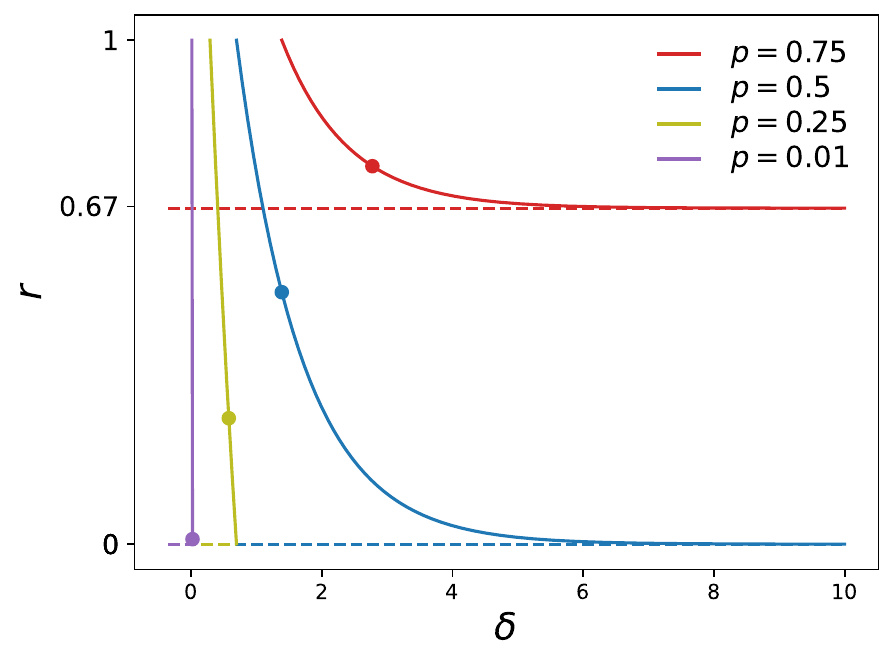}
    \includegraphics[width=0.39\textwidth]{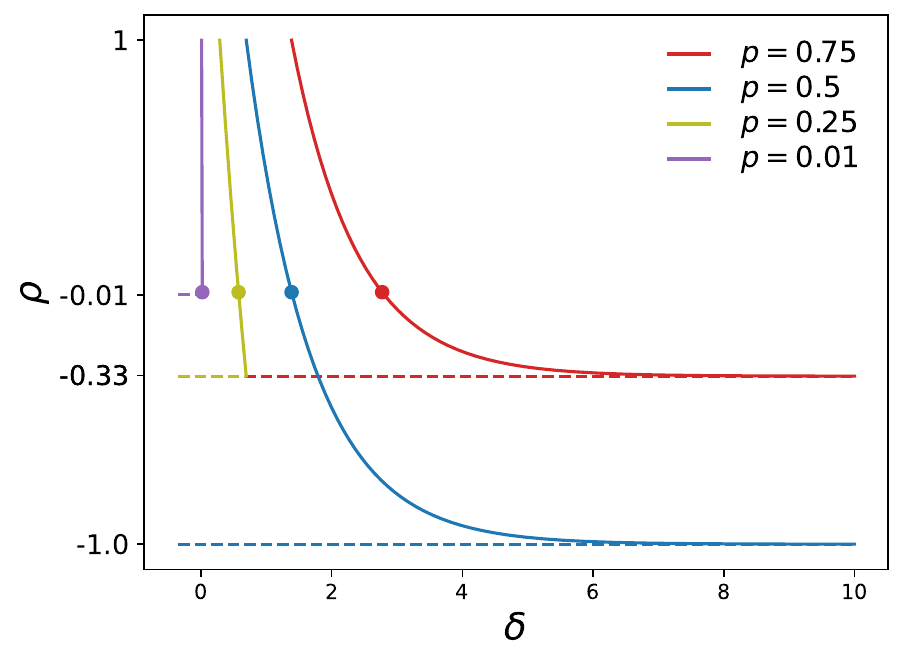}
    \caption{\small\textbf{Reciprocity ($r$) and  correlation coefficient ($\rho$) as functions of the density parameter ($\delta$)}. The solid lines illustrate, for different values of the link density $p$, the behavior of the functions $\langle r\rangle$) and  $\langle\rho\rangle$ in equations (\ref{eq:rho}), while the isolated points indicate the values achieved in the areciprocal case $\delta_{\text{rand}} = 2 \epsilon$ (where $\langle r \rangle_{\text{rand}} = p$ and $ \langle\rho\rangle_{\text{rand}} = 0$). The dashed lines indicate the minimum values of $\langle r\rangle$ and $\langle \rho\rangle$ attainable for the given values of $p$.
    The four values of $p$ considered are: 1) $p=0.75$, for which the maximum value $q_\text{max}$ for $q$ is $1$, $\langle r\rangle_{\min} = \frac{2p - 1}{1-p} $ and $\langle \rho\rangle_{\min} =  \frac{2 -p-1/p}{1-p} $; 2) $p=0.5$, for which $q_{\max} = 1$, $\langle r\rangle_{\min} = 0 $, and $\langle \rho\rangle_{\min} = - \frac{p}{1-p} $; 3) $p=0.25$, for which $q_{\max} = 0.5$, $\langle r\rangle_{\min} = 0 $, and $\langle \rho\rangle_{\min} =  -\frac{p}{1-p} $; 4) $p= 0.01$, for which $q_{\max} \sim q_{\text{rand}}$, $\langle r\rangle_{\min} = 0 $, and $\langle \rho\rangle_{\min} =  -\frac{p}{1-p} \sim 0$.}
    \label{fig: ER}
\end{figure}

We are mainly interested in the amount of reciprocation generated by the model, compared with an areciprocal benchmark, or equivalently with the random reciprocity $\langle r \rangle_\text{rand} = p $ generated by model itself in the particular case when $p^\leftrightarrow=p^2$, i.e. for $\delta_\text{rand}=2\epsilon$ (this case is simply a directed Erd\H{o}s-R\'enyi model).
Discounting for this `baseline' reciprocity can be done transparently by adopting a modified metric of reciprocity \cite{Garlaschelli2004PatternsOL} defined as the Pearson correlation coefficient $\rho$ between the symmetric entries of the adjacency matrix, which for homogeneous random graphs takes the convenient form
\begin{equation*}
    \rho \equiv \frac{\sum_{i, j\neq i} (a_{ij} - \langle a_{ij} \rangle)(a_{ji} - \langle a_{ij} \rangle)}{\sum_{i, j\neq i}(a_{ij} - \langle a_{ij} \rangle)^2} = \frac{ r   -  p  }{1- p  }.
\end{equation*}
Unlike $r$, $\rho$ is an absolute quantity and takes values in the interval~$[-1,1]$. 
The value $\rho = 0$ points at the areciprocal case, where the couples $a_{ij}$ and $a_{ji}$ are indeed statistically independent.
The sign of $\rho$ immediately distinguishes then between reciprocal ($\rho > 0$) and antireciprocal ($\rho < 0$) networks.
The explicit expressions for the expectations of $r$ and $\rho$ are
\begin{equation} \label{eq:rho}
    \begin{split}
        & \langle r \rangle  = 2 - \frac{q}{p} = 2 - \frac{1-e^{-\delta}}{1-e^{-\epsilon}},\\
        & \langle \rho \rangle  = \frac{(2-p)p-q}{(1-p)p} = \frac{ e^{-(\delta - \epsilon)} - e^{- \epsilon}}{1-e^{- \epsilon}}.
    \end{split}
\end{equation}
The above expressions are illustrated in  Fig.~\ref{fig: ER}, where the dependence of $\langle r \rangle$ and $\langle \rho \rangle$ on $\delta$ is shown for different values of $\epsilon$ (i.e. of the expected link density $p$). 
It is interesting to note how the antireciprocal regime converges to the areciprocal one as $\epsilon\to 0$: in this limit, the graph gets so sparse that it becomes extremely unlikely for two nodes to establish mutual connections, and reciprocated links are effectively absent.
This can be confirmed by expanding the conditions in~\eqref{eq: CE} around $\epsilon=0$: the value of $\delta$ for which $p^{\leftrightarrow}$ vanishes  is $ \delta_{\max}   \stackrel{\epsilon \sim 0}{\sim} - \ln(1-2\epsilon) \sim 2 \epsilon $, which coincides with the value of $\delta$ for which $p^{\leftrightarrow} = p^2$ (i.e. the areciprocal case). Note that, starting from such a sparse configuration at level $\ell=0$ and progressively coarse-graining the network implies a departure from this scenario, as reciprocated links become more and more likely.

\begin{figure*}[t!] 
     \centering
     \includegraphics[width=0.32\textwidth]{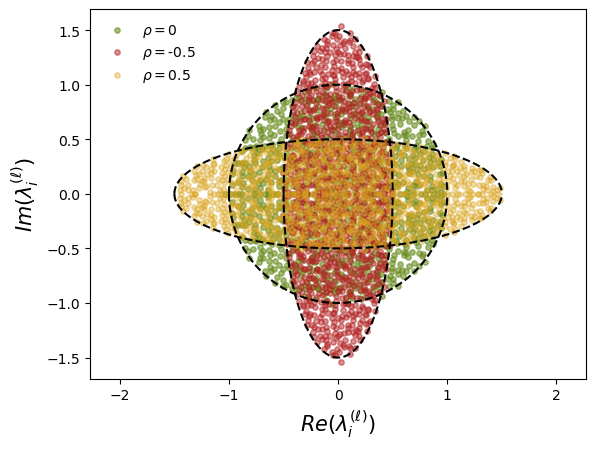}
     \includegraphics[width=0.32\textwidth]{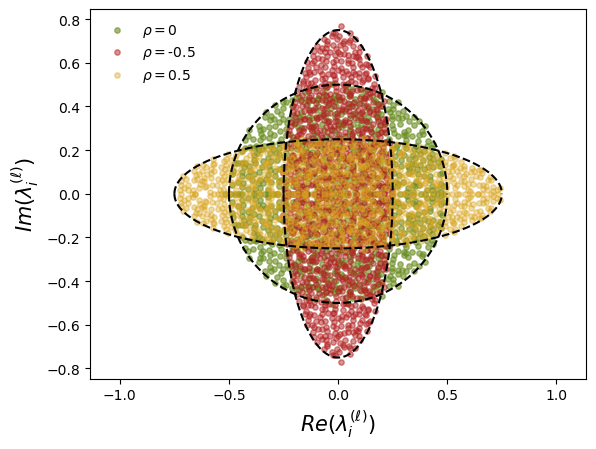}
     \includegraphics[width=0.32\textwidth]{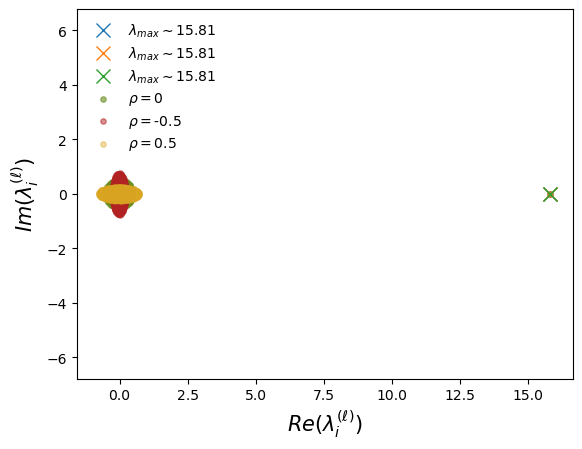}\\
     \includegraphics[width=0.32\textwidth]{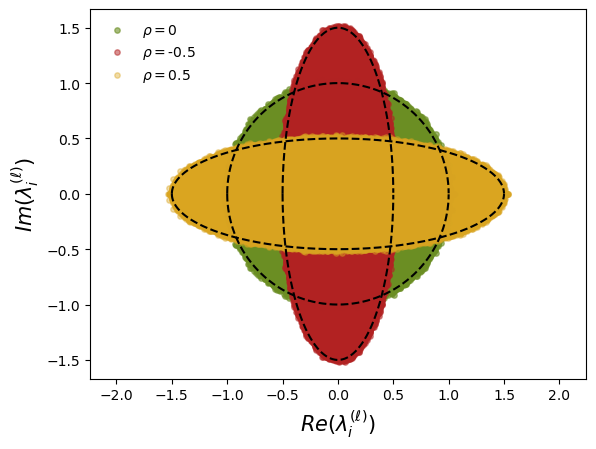}
     \includegraphics[width=0.32\textwidth]{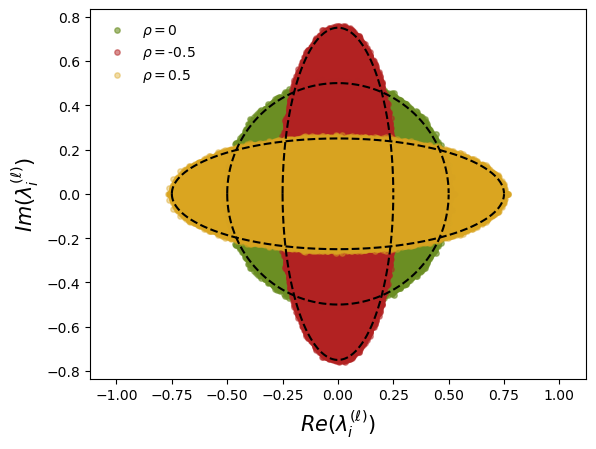}
     \includegraphics[width=0.32\textwidth]{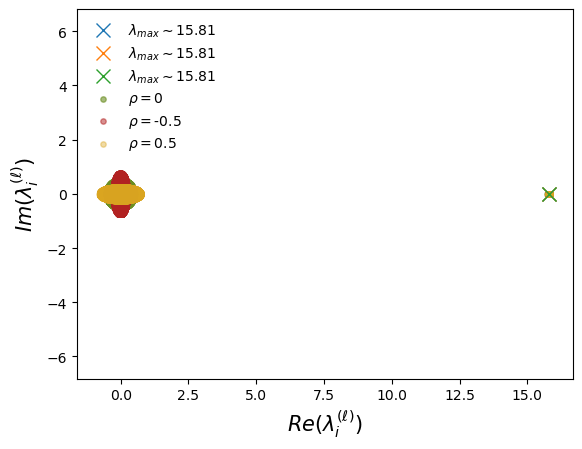}
     \caption{\small \textbf{Spectral density for homogeneous directed networks}. Complex spectra of the (centered and rescaled) adjacency matrix of the network for 1 (top panels) and 100 (bottom panels) realizations of the homogeneous DSIM with $N=1000$ nodes. In all cases the parameter $\epsilon$ is kept fixed to the value $\epsilon  = 0.69315$, corresponding to the link density $d= 0.5$, which allows the parameter $\rho$ to vary across its full range $[-1,1]$ (see Fig.~\ref{fig: ER}). The spectra are shown for the three values  $\delta = 2.0794$ ($\rho = -0.5$, red), $\delta =2\epsilon$ ($\rho = 0$, green), and $\delta =0.9808$ ($\rho = 0.5$, yellow). Right: spectra of the matrix  with entries $C_{ij} = \frac{a_{ij}-p}{\sqrt{N p(1-p)}}$, which are uniform over the $\text{Ellipse}[1+\rho, 1-\rho]$. 
     Center: spectra of the matrix with entries $\frac{a_{ij}-p}{\sqrt{N}}$, which are uniform over the $\text{Ellipse}[(1+\rho)\sqrt{p(1-p)},(1-\rho)\sqrt{p(1-p)}]$. 
     Right: spectra of the matrix with entries $a_{ij}$, which consist of an elliptic `bulk' and an isolated largest real eigenvalue $\lambda_\text{max}$ (crosses). Note that $\lambda_\text{max} \sim N p$ does not depend on $\delta$ and on the resulting level of reciprocity. }
    \label{fig:spectral ER}
 \end{figure*}

A remarkable effect of reciprocity is visible in the eccentricity of the elliptical spectral density of the adjacency matrix of directed graphs in the complex plane, as pointed out in~\cite{sommers1988spectrum}.
In our setting we can define the centered random matrix $\mathbf{C}$ with elements  $C_{ij} = \frac{a_{ij} - p_{ij}}{\sqrt{Np_{ij}(1-p_{ij})}}$, which have zero mean and unit variance. We can easily write the covariance between its off-diagonal entries in the suggestive form
\begin{equation*}
     \langle C_{ij} C_{ji} \rangle = \frac{p^{\leftrightarrow} - p^2}{N p(1-p)} = \rho\quad i\ne j.
 \end{equation*}
According to~\cite{sommers1988spectrum}, the spectral density $\sigma$ of $\mathbf{C}$ is (asymptotically) uniform on the ellipse with axes $(1 + \rho, 1- \rho)$.
In terms of reciprocity, this statement has an intuitive meaning: the larger the value of $\rho$ (i.e. the `more symmetrical' the matrix $\mathbf{C}$), the more the spectral ellipse `narrows' along the real axis, until it reaches the undirected (fully symmetric) limit, for which $\rho = 1$ and the Wigner semicircle law is recovered~\cite{wigner1958distribution}. On the opposite, the smaller is $\rho$, the more the ellipse concentrates on the imaginary axis.
In the intermediate areciprocal case, we find a perfect circle ($1 + \rho=1- \rho=1$) as expected for matrices with perfectly uncorrelated entries~\cite{girko1986elliptic}. 
The above picture is  illustrated in Fig.~\ref{fig:spectral ER}, where the outcome of numerical simulations is shown either for a single realization of the homogeneous model and for a sample of 100 realizations.
Note that having centered the matrix $\mathbf{A}$ effectively removes the largest eigenvalue $\lambda_{\max}(\mathbf{A})$ (which is known to behave as $ \sim Np$ for Erd\H{o}s-R\'enyi random graphs, either directed o undirected) from the spectrum of $\mathbf{C}$, as illustrated in the right panels in Fig.~\ref{fig:spectral ER}. Moreover, rescaling by $\sqrt{Np_{ij}(1-p_{ij})}$ is convenient as $N$ gets large. 
 
The remarkable consistency between individual realizations and the theoretical elliptic distribution, already for moderate network sizes, justifies the use of the spectral density of the adjacency matrix of (necessarily finite) real-world networks as a powerful quantification of reciprocity, as we will see in the next Section on a specific case study.

\section{The quenched scenario: modelling real-world networks}\label{Sec:Directed-quenched}
In this section, we move away from the homogeneous setting and address the general case where different nodes can have different fitness values, while still treating the latter as deterministic variables (which therefore represent some sort of `intrinsic characteristics' of nodes).
In particular, each node $i_{\ell}$ is provided with a fitness quadruplet $(x_{i_{\ell}}, y_{i_{\ell}}, z_{i_{\ell}}, w_{i_\ell})$ that will determine its tendency of establishing directed links with other nodes and with itself. When using the model to describe a real-world system,  every element of the latter may be represented as a node at a certain level $\ell$, while the fitness parameters may be identified with measurable node-specific quantities. 
Note that we may choose to tune only the total number of self-loops and not their individual node-specific probabilities, thereby relying solely on the parameter $\eta$ and fixing $ w_{i_\ell}=1 \; \forall i_\ell = 1,\dots N_\ell$.
As already mentioned, we may also add a  (symmetric) dyadic property $d_{i_{\ell}, j_{\ell}}$ to each pair of nodes $(i_{\ell}, j_{\ell})$, to further encourage (or discourage) their connection.
In this way, the only remaining free parameters are the global, scale-invariant ones ($\epsilon$, $\delta$, $\eta$). The latter can be adjusted by constraining the link density, the global reciprocity, and the overall number of self-loops, respectively.
Once the parameters are specified, the connection probabilities in~\eqref{pright}-(\ref{pboth}) are determined simultaneously at all scales, thereby providing a testable multiscale model of the renormalized network at any higher level of aggregation.
In the following Subsection, we illustrate the procedure by considering the empirical International Trade Network (ITN).

\subsection{Modelling the International Trade Network}
In~\cite{Elenulla} it was shown how the undirected SI model can reproduce the topology of the observed (symmetric) network of bilateral international trade, not only at the `native' level of resolution (where nodes represent countries and links represent the existence of trade in any direction) but also consistently across different aggregation levels, e.g. between groups of countries, regions, etc.. 
Here we walk the same line while focusing on the finer topology of \emph{directed} international trade (where a directed link represents export from one node to another). This is not a straightforward extension of the undirected case, precisely because of the strongly reciprocal nature of this particular directed network, which has been pointed out for example in \cite{Garlaschelli2004PatternsOL, garlaschelli2005structure}. Capturing the reciprocity structure of the ITN is crucial, among other things, for correctly assessing the statistical significance of higher-order patterns such as triadic motifs~\cite{squartini2012triadic}.
We consider data from the expanded trade dataset developed by K. S. Gleditsch~\cite{gleditsch2002expanded}, which provides trade flow estimates among  $185$ countries for the years $1948-2000$. 
We are going to show the results for the year 2000, but we have obtained similar outcomes for the other years as well (more details on the dataset are provided in SI).

We define our empirical network at the native level $\ell = 0$ by considering $N_0=185$ 0-nodes, each representing a country in the dataset \cite{gleditsch2002expanded}. The empirical adjacency matrix $\Tilde{\mathbf{A}}^{(0)}$ (where the tilde indicates the specific empirical matrix in the ensemble of possible matrices that the model we are going to define can generate) is then built by setting the entry $\Tilde{a}^{(0)}_{i_{0} j_{0}} = 1$ if there is any reported (i.e. positive) export relationship from country  $i_{0}$ to the country $j_{0}$, and $\Tilde{a}^{(0)}_{i_{0} j_{0} } = 0$ otherwise.
We set by convention all self-loops to zero: $\Tilde{a}^{(0)}_{i_{0} i_{0} } = 0, \; \forall \, i_{0}=1,\dots,N_0$. 
In what follows, we define our multiscale model for the directed ITN by using the probabilities introduced in~\eqref{pright}-(\ref{pboth}) and specifying all the parameters therein.

We first discuss the choice of the fitness variables $x_{i_0}$ for each country $i_0 = 1,N_0$.
Building upon prior findings that highlight the significance of the Gross Domestic Product (GDP) in shaping the structure of the undirected ITN~\cite{garlaschelli2004fitness,almog2015gdp,almog2017double, almog2019enhanced}, we follow the approach taken in~\cite{Elenulla} and set $x_{i_0}=\text{GDP}_{i_0}$ for all $i_0=1,N_0$, where the GDP is sourced from the Gleditsch database itself~\cite{gleditsch2002expanded}. 
Notably, this choice is consistent by the additive nature of GDP upon aggregation (the total GDP of a set of countries is the sum of the individual GDPs), which compels with the renormalization rule for $x$ in~\eqref{renormxw}. 

Next, we fix the dyadic parameters, a natural candidate for which is in this case provided by the geographic distances between countries.
Again, this builds upon established empirical knowledge of the role of distances in determining trade flows, as initially conceptualized by the traditional Gravity Model of trade~\cite{tinbergen1962shaping} and more recently confirmed by various network analyses of the ITN~\cite{almog2019enhanced,di2022gravity,di2023reconciling}.
Following the undirected analysis in~\cite{Elenulla}, we therefore 
consider the population-averaged inter-country distances $\{d_{i_0 j_0}\}$ provided by the  BACI-CEPII GeoDist database~\cite{mayer2011notes,head2002illusory} (see SI).
However, as discussed in general in Sec.~\ref{section: DSIM}, we need our distances to be \emph{metaultrametric} over the dendrogram induced by the desired sequence $\{\Omega_\ell\}_{\ell\ge 0}$ of node partitions. 
For this reason, for all $i_0$ we set our diagonal metaultrametric distances $d^\text{mu}_{i_0 i_0}$ equal to the non-zero self-distances $d_{i_0 i_0}>0$ as already reported in the GeoDist data (representing the population-averaged distance between agglomerations inside each country). Next, we construct the off-diagonal metaultrametric distances $d^\text{mu}_{i_0 j_0}$ (for $i_0\ne j_0$) via a single-linkage hierarchical clustering algorithm, which produces the so-called \emph{subdominant ultrametric} distances representing the closest ultrametric `from below' to the original metric~\cite{rammal1986ultrametricity} (see SI for details).
The subdominant ultrametric distances are obtained by using the original geographic distances $\{d_{i_0 j_0}\}$ as measure of dissimilarity.  
The output of this procedure is a dendrogram (shown in SI) whose leaves correspond to the original countries (0-nodes) and the metaultrametric distance between each pair is given by the height of the branching point for the corresponding branches. 
The dendrogram will then be used to induce the nested partitions of nodes in the coarse-graining procedure, by applying a sequence of `horizontal cuts'. 
Note that, at any hierarchical level $\ell$, the number of different values of the distances between $\ell$-nodes is $N_\ell-1$ (which equals the number of branching points of the dendrogram above the horizontal cut), plus the $N_\ell$ self-distances.
Then, following~\cite{Elenulla}, we set the function $f(d^\text{mu} )\equiv {1}/{d^\text{mu} }$, in analogy with how in the GeoDist database the population-averaged (here, GDP-averaged) distances between countries are calculated according to a formula similar to~\eqref{renormfxx}.

We then consider the fitness $w_{i_0}$. To replicate the absence of empirical self-loops at level $0$ we set $w_{i_0} \equiv  x_{i_0}^2 f(d^\text{mu}_{i_0 i_0})$ for all $i_0=1,N_0$ and $\eta \equiv-{\delta}/{2}$ (the latter is analogous to the choice made in Sec.~\ref{section: ER}).
Note that, as we proceed to higher levels $\ell>0$ through coarse-graining, self-loops will eventually emerge both in the empirical network and in the model.

Finally, we consider the fitness values $y_{i_0}$ and $z_{i_0}$.
As already observed in \cite{Garlaschelli2004PatternsOL} and confirmed for this specific dataset (see SI), each country in the ITN exhibits approximately equal in-degree $k_{in}(i_{0}) = \sum_{j_0\neq i_0} a_{j_0 i_0}$ and out-degree $k_{out}(i_{0}) = \sum_{j_0\neq i_0} a_{i_0 j_0}$. 
As a consequence, for this particular system, a natural choice is the symmetric one $p_{i_0j_0} \equiv p_{j_0i_0}$, i.e. $y_{i_0}\equiv z_{i_0}$, $\forall \, i_0,j_0$.
Moreover, to keep the focus on GDP alone as the main node-specific driver of trade links, we make the convenient choice of taking the vectors $\vec{y}$ and $\vec{z}$ proportional to $\vec{x}$. Since the (positive) proportionality constant can be reabsorbed in the global parameter $\epsilon$, we posit $\vec{x}\equiv\vec{y}\equiv\vec{z}$.

With the above specifications, the only free parameters left are $\epsilon$ and $\delta$, which we tune, respectively, so that both the expected number $\langle L^\text{nl}_0\rangle$ of directed links (self-loops excluded) and the expected number $\langle L^{\leftrightarrow}_0\rangle$ of reciprocated links equal their measured counterparts $\tilde{L}^\text{nl}_0$ and $\tilde{L}^{\leftrightarrow}_0$ in the empirical network at level $\ell=0$ (see SI).

\subsection{Geographical aggregations}
We can now test the model against the real data at several levels of aggregation, through a sequence of iterative coarse-grainings induced by the dendrogram of metaultrametric distances discussed above.
We follow the same procedure adopted in~\cite{Elenulla} and cut the dendrogram horizontally at 17 different levels, such that the number of block-countries, at any coarse-graining level $\ell$, is given by $N_0 = 185$ for $\ell = 0$ and $N_{\ell} = 180 - 10\ell, \, \forall \; \ell = 1,\dots , 16$.
This procedure generates the hierarchy of coarse-grained empirical networks $\{\Tilde{\mathbf{A}}^{(\ell)} \}_{\ell = 1, \dots, 17}$ via the procedure illustrated in Fig.~\ref{fig:toyCG}.
In this way, each $\ell$-node represents a geographic aggregate of nearby countries, and the  renormalized fitness represents the empirical aggregate GDP of a block-country.
Moreover, the transformation rules in~\eqref{renormfxx} and~\eqref{renormfyz} leave the off-diagonal metaultrametric distances $f(d^\text{mu}_{i_{{\ell}+1},j_{{\ell}+1}})$ (for $i_{{\ell}+1}\ne j_{{\ell}+1}$) invariant at any level $\ell$, as specified in~\eqref{eq:SIf}, while the diagonal terms $f(d^\text{mu}_{i_{{\ell}+1},i_{{\ell}+1}})$ renormalize as in~\eqref{renormfxx}.
The values of the model parameters, together with their transformation rules across hierarchical levels, determine the connection probabilities simultaneously at all scales considered.
We have checked that the conditions in~\eqref{eq: CE} are verified along the entire coarse-graining flow.

\begin{figure*}[t]
    \centering
    \includegraphics[width=0.32\textwidth]{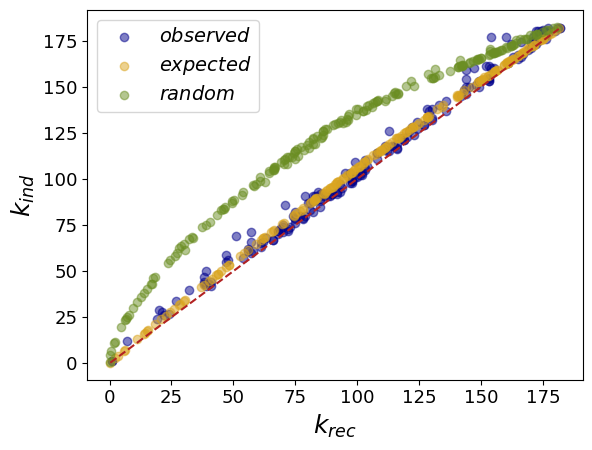}%
    \includegraphics[width=0.32\textwidth]{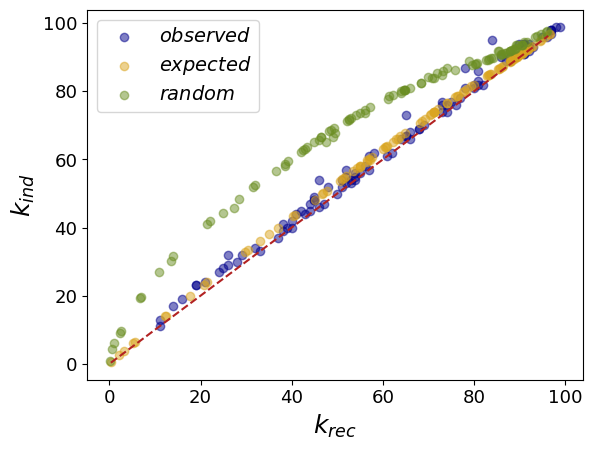}%
    \includegraphics[width=0.32\textwidth]{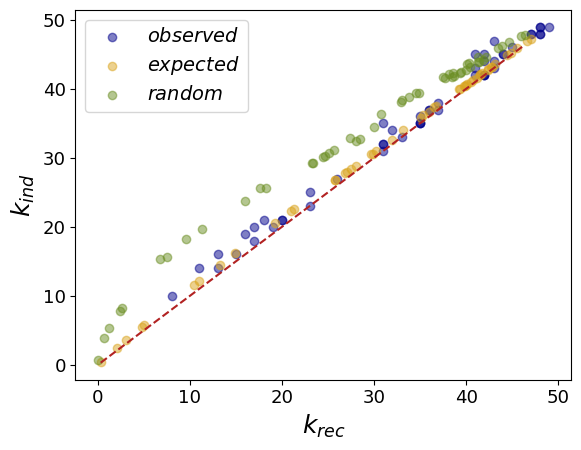}
    \includegraphics[width=0.32\textwidth]{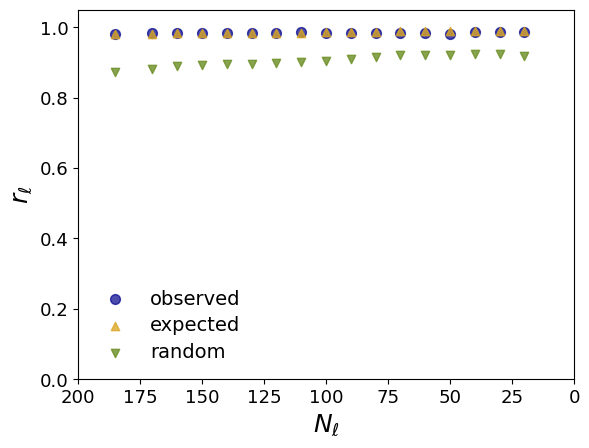}%
    \includegraphics[width=0.32\textwidth]{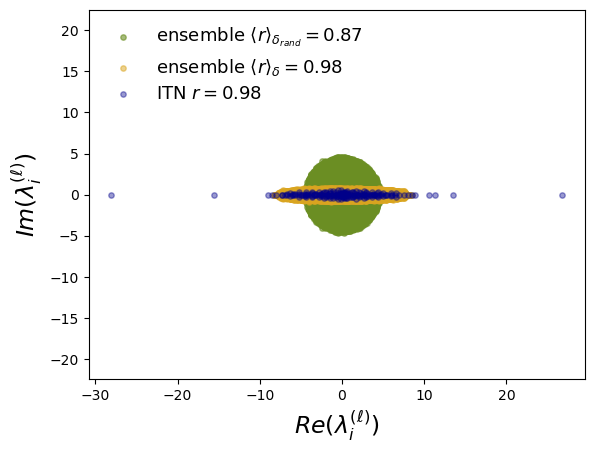}%
    \includegraphics[width=0.32\textwidth]{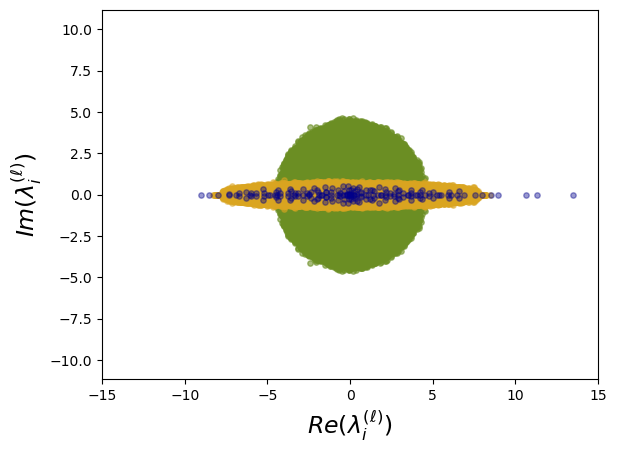}
    \caption{\small\textbf{Topological and spectral properties of the ITN across different levels of aggregation}.  
    Various features of the empirical ITN (in blue) are compared with the expectations provided by the DSIM (in yellow) and by the \textit{areciprocal DSIM} - where $\delta = \delta_{\text{rand}}$ (in green).
    Top panels:  scatter-plot of the undirected projection of the degree $k^{\text{und}}_{i_{\ell}}$ vs the reciprocal degree $k^{\leftrightarrow}_{i_\ell}$ of each node, at three representative levels of aggregation:  starting from $\ell=0$ with $N_0 = 185$ nodes, corresponding to the original ITN (on the left), to  $\ell=8$ with $N_8 = 100$ nodes (at the centre) to  $\ell=13$ with $N_{13} = 50$ nodes (on the right).
    Bottom panels: in the figure on the left, the observed and expected values of global reciprocity $r_\ell$ are shown as functions of the number  $N_\ell$ of nodes along the $17$ considered levels of aggregation of the original network, with $N_0 = 185$ and $N_\ell = 180 - \ell \, 10$ for any $\ell=1,\dots 16$.
    In the two figures on the right, we focus on the spectral density of the ITN and show real and imaginary parts of the eigenvalues of the empirical adjacency matrix of the ITN (in blue), of $N_g=1000$ adjacency matrices drawn from the DSIM ensemble (with the same density and same reciprocity of the empirical network) (in yellow) and of $N_g=1000$ adjacency matrices drawn from the areciprocal DSIM (with the same density as the empirical network, but random reciprocity) (in green). For visualization purposes, in all cases the spectrum of the centred matrix $\mathbf{A}_c$ (with elements  $a^{(c)}_{ij} = a_{ij} - p_{ij}$) is considered instead of the one of the adjacency matrix $\mathbf{A}$. The picture on the right offers a soft zoom on the bulk of the spectrum shown on the left.}
    \label{fig:ITN-rec}
\end{figure*}
\subsection{Capturing structural and spectral properties}
As anticipated, we find that the strongly reciprocated nature of the ITN cannot be captured without explicitly enforcing it, since the empirical global reciprocity in the network is not consequential to other topological properties.
This can be easily seen by comparing various features of the real ITN with the theoretical expectations provided by our DSIM in two different settings: a \textit{fitted} one, where the parameter $\delta$ is tuned to replicate the empirical number of reciprocated links (as explained above and detailed in SI), and a \textit{random} one, where $\delta_{\text{rand}} = \frac{\epsilon(y_i z_j + y_j z_i)}{x_i x_j}$ as in the areciprocal benchmark for our model discussed in~\ref{parag:relevant}.
In the symmetric ($y_i=z_i$) setting considered here, this is simply obtained as  $\delta_{\text{rand}} = 2 \epsilon$. 
We consider both topological and spectral properties, for various hierarchical levels: specifically, the global reciprocity $r_\ell$, the reciprocated degree $k^{\leftrightarrow}_{i_{\ell}}\equiv\sum_{j_{\ell}\ne i_\ell}a_{i_\ell j_\ell}a_{j_\ell, i_\ell}$ of each node $i_{\ell}$ and the spectral density of the adjacency matrix ${\mathbf{A}}^{(\ell)}$. 
The results are shown in Fig.~\ref{fig:ITN-rec}.

First, we note from the bottom left panel of Fig.~\ref{fig:ITN-rec} that the global reciprocity $r_\ell$ is always systematically larger in the empirical network than in the areciprocal model with the same link density. By contrast, the reciprocated model replicates $r_0$ by construction, via the fitted parameter $\delta$. Yet, it is quite remarkable that the agreement between the empirical and the predicted reciprocity remains in place across all subsequent aggregation levels $\ell>0$, even if no further refitting of the parameters has taken place.

Similar considerations apply to the relationship between the reciprocal degree  $k^{\leftrightarrow}_{i_{\ell}}$ and the undirected degree $k^{\text{und}}_{i_{\ell}} $, which is shown for
three representative levels of aggregation (top panels of Fig.~\ref{fig:ITN-rec}).
Note that the areciprocal directed model cannot replicate the empirical (approximately linear) relationship between the two quantities, the reciprocated degree being systematically underestimated. By contrast, the full model replicates the empirical replationship remarkably well: even if only the expected global reciprocity at level $\ell=0$ is fitted through the parameter $\delta$, the model successfully predicts the local reciprocated degree for each node separately, and keeps doing so (again, without refitting) across aggregation levels. 

Finally, we consider the spectral properties of the network, in analogy with the analysis in Sec.~\ref{section: ER}, but here for heterogeneous fitness values. 
We compare the spectral distribution of the ITN with the one of samples drawn from the DSIM ensemble.
We generated $N_g = 1000$ synthetic realizations of the empirical ITN for each of the two models considered (with $\delta$ fitted and $\delta_{rand}$) and then juxtaposed their spectra with the empirical one. 
The outcome is shown on the bottom right of Fig.~\ref{fig:ITN-rec}, where the strongly eccentric (along the real axis) elliptical distribution of the ITN is closely replicated by the model with reciprocation, while the areciprocal model produces the usual circular distribution. 

To the best of our knowledge, what we presented in this Section is the first directed model of the ITN that replicates its reciprocity and spectral properties in a way that remains consistent across arbitrary geographic aggregations. The model succeeds in replicating the empirical properties at all aggregation levels by fitting only the two global parameters $\epsilon$ and $\delta$, and only at level $\ell=0$.

\section{The annealed scenario: emergent reciprocal networks} \label{Sec:Directed-annealed}
As in the undirected case~\cite{Elenulla}, the DSIM can be framed in an annealed scenario, where not only the graph structure but also the fitness values are treated as random variables.
Then, the requirement of invariance upon aggregation is extended to the functional form of the probability density function (pdf) describing these random variables. 
In the directed case, each node is in general endowed with four random fitness variables, so at each level $\ell$ there are four $N_\ell$-dimensional random vectors ($\vec{x}^{(\ell)}, \vec{y}^{(\ell)}$, $\vec{z}^{(\ell)}$ and $\vec{w}^{(\ell)}$).

\subsection{Model specification}
For simplicity, we disregard the dyadic factors, i.e. we posit $f\equiv 1$ in Eqs.~\eqref{pright},~\eqref{pleft},~\eqref{pno},~\eqref{pboth}, and leave to chance the formation of self-loops by setting $\vec{w}^{(0)}\equiv\vec{0}$.
The additivity of $w$ upon aggregation then also implies $\vec{w}^{(\ell)}=\vec{0}$ deterministically for all higher levels $\ell>0$.
We are then left with only three random vectors ($\vec{x}^{(\ell)}, \vec{y}^{(\ell)}$ and $\vec{z}^{(\ell)}$) which can in general be mutually dependent in complicated ways, both across nodes (e.g. $x_{i_\ell}$ and $x_{j_\ell}$ can be correlated) and across fitness types (e.g. $x_{i_\ell}$, $y_{i_\ell}$ and $z_{i_\ell}$ can be correlated. 
While it is reasonable to assume statistical independence between different nodes (in the general spirit of fitness models or inhomogeneous random graphs considered in the literature), doing the same between different fitness types would lead to an unrealistic scenario where, for instance, there would be no correlation between the in-degree and out-degree of the same node (while, as the ITN example considered above illustrates in a particularly strong way, these variables are in general correlated).
For the sake of simplicity, we consider the extreme case of perfect correlation by setting $\vec{x}^{(\ell)}\equiv \vec{y}^{(\ell)}\equiv\vec{z}^{(\ell)}$ for all $\ell\ge 0$. 

With the above choices, the multivariate problem reduces to a univariate one, where each $\ell$-node $i_\ell$ is endowed with a single random fitness $x_{i_\ell}$ (the fitness values for all nodes being \emph{i.i.d.}) that impacts all its types of connections through the probabilities defined by Eqs.~(\ref{pright})-(\ref{pboth}), which here become
\begin{equation} \label{Deq:annealed p}
    \begin{split}
        &p_{i_\ell j_\ell} = 1 - e^{-\epsilon x_{i_\ell} x_{j_\ell}}, \\
        &q_{i_\ell j_\ell} = 1 - e^{-\delta x_{i_\ell} x_{j_\ell}},\\
        &p_{i_\ell j_\ell}^{\rightarrow}  = p_{i_\ell j_\ell}^{\leftarrow} = e^{-\epsilon x_{i_\ell} x_{j_\ell}} - e^{-\delta x_{i_\ell} x_{j_\ell}},\\
        &p_{i_\ell j_\ell}^{\leftrightarrow}  = 1 +e^{-\delta x_{i_\ell} x_{j_\ell}} -2e^{-\epsilon x_{i_\ell} x_{j_\ell}},
    \end{split}
\end{equation}
for any $i_\ell \neq j_\ell$, while the diagonal terms read $p_{i_\ell i_\ell} = q_{i_\ell i_\ell} =  p_{i_\ell i_\ell}^{\leftrightarrow} = 1 - e^{-\frac{\delta}{2} x_{i_\ell}^2}$.\\

As discussed in~\cite{Elenulla}, requiring that the pdf of the (necessarily positive) fitness values is positively supported and invariant upon aggregation (just like the graph probability distribution) leads to the selection of \emph{one-sided $\alpha$-stable distributions}~\cite{penson2010exact}. In this way,
at any level $\ell$ of observation, a realization $\{x_{i_\ell}\}_{i_\ell =1}^{N_\ell}$ of fitness variables can be obtained in two equivalent ways, either \emph{hierarchically} from any lower hierarchical level (say $\ell -1$) by summing the finer-grained variables $\{x_{i_{\ell-1}}\}_{i_{\ell-1} =1}^{N_{\ell-1}}$ based on the chosen partition $\Omega_{\ell-1}$, or \emph{directly} by drawing $N_{\ell}$ i.i.d. random variables from a pdf with invariant functional form and scale-dependent parameters. 
The aforementioned property is ensured by the \emph{closure under convolution} that characterizes $\alpha$-stable random variables.
We denote the pdf of $\alpha$-stable distributions as $\varphi_\ell(x_{i_\ell}|\alpha, \beta_\ell, \gamma_\ell, \mu_\ell)$, where $\alpha$ is the stability parameter (which, in the one-sided case, ranges in the interval $0<\alpha<1$), $\beta_\ell$ controls the skewness, $\gamma_{\ell}$  the scale and $\mu_\ell$ the location of the distribution. 
In the specific case of one-sided $\alpha$-stable distributions with support over the non-negative real numbers, we have $\mu_0 = 0$ and $\beta_0=1$.
Upon aggregation, the only  parameter that gets modified is $\gamma_{\ell}$, while $\alpha$ remains unchanged due to the stability property and $\beta_\ell$ and $\mu_\ell$ are mapped onto their original values $\beta_0=1$ and $\mu_0=0$.
In order to preserve the \emph{i.i.d.} nature of the fitness across $\ell$-nodes, we assume that, to pass from level $\ell$ to $\ell+1$, blocks of homogeneous sizes are formed, i.e. all $(\ell+1)$-nodes contain exactly the same number (say, $b_\ell$) of $\ell$-nodes.
Under this assumption, $\gamma_{\ell}$ transforms as $\gamma_{\ell+1}^{\alpha} = b_\ell \gamma_{\ell}^{\alpha}$.

In the remainder of this section, we narrow the focus on the particular case of the L\'{e}vy distribution, corresponding to $\alpha=1/2$ and representing the only $\alpha$-stable distribution that can be written in explicit form when $0<\alpha<1$. The L\'{e}vy pdf reads
\begin{equation} \label{eq: Levy distr}
    \varphi(x | \gamma_{i_{\ell}})\equiv \varphi_\ell(x | {1}/{2},  1, \gamma_{i_{\ell}}, 0) = \sqrt{\frac{\gamma_{i_{\ell}}}{2 \pi}} \frac{e^{-{\gamma_{i_{\ell}}}/{2 x}}}{x^{3/2}}.
\end{equation}
Note that, as for any $\alpha$-stable distribution with $0<\alpha<1$, all moments of the above distribution diverge. In particular, the divergence of the mean crucially changes the picture with respect to most other inhomogeneous random graph models~\cite{avena2022inhomogeneous}, which typically have finite-mean fitness.

A further, crucial property of stable distributions is  \emph{infinite divisibility},  i.e. they can always be expressed as the probability distribution of the sum of an arbitrary number of \emph{i.i.d.} random variables from the same family. 
This implies that we can disaggregate each node at level $\ell$ with fitness $x_{i_\ell}$ into any desired number of nodes at level $\ell-1$, each with its own fitness, thereby introducing a consistent fine-graining procedure in which, in principle, the graph $G_{N_0}$ can resolved into a finer graph $G_{N_{-\ell}}$ (at a `negative' hierarchical level $-\ell$) with any number of nodes $N_{-\ell}>N_0$. Consequently, in the annealed case, the renormalization procedure establishes \emph{a group structure}, where both the bottom-up scheme (proceeding from lower to higher hierarchical levels) and the top-down scheme (proceeding from higher to lower levels) are consistently defined.

In what follows, we explore the topology of network realizations drawn from the DSIM ensemble in the annealed setting described above, i.e. by connecting pairs of $\ell$-nodes with the connection probabilities given in Eq.~\eqref{Deq:annealed p} and L\'{e}vy distributed fitness variables.
In particular, we investigate how the parameters $\epsilon$ and $\delta$ jointly  shape the features of such networks at different levels of aggregation.

\subsection{Emergent positive reciprocity}
Before delving into a multiscale analysis, we first note that, under our choice of L\'evy-distributed fitness and $\vec{x}^{(\ell)}\equiv \vec{y}^{(\ell)}\equiv\vec{z}^{(\ell)}$, the formation of antireciprocal networks is suppressed.
To support the former statement we recall from the discussion of the homogeneous case in Sec.~\ref{section: ER}  that finding the value $\delta_{\max}$ such that $\langle r \rangle_{\delta_{\max}} = 0$ (corresponding to the maximally antireciprocal limit) is possible only as long as the overall link density (i.e. $p$) remains below the structural limit of $1/2$ (recall Fig.~\ref{fig: ER}).
At the same time, as mentioned in Sec.~\ref{parag:relevant}, strong heterogeneity in the fitness distribution makes it very difficult for the resulting network to exhibit antireciprocal features, even when the overall density of links is very low.
Indeed, the upper bound $\delta_{\max}$ in Eq.~\eqref{eq: CE} is dictated by the pair of nodes with the lowest fitness as $\delta_{\max} = \min_{i,j}{ \delta_{i_0j_0}^{\max} }$, where $\delta_{i_0 j_0}^{\max} = - \frac{\ln(2e^{-\epsilon x_{i_0} x_{j_0}} -1)}{x_{i_0} x_{j_0}}$ is a strictly increasing function of $x_{i_0} x_{j_0}$. As a consequence, the wider the gap between the minimum and maximum fitness products in a given realization, the broader the range of values of $\delta_{i_0j_0}^{\max}$ across pairs of nodes and the larger the number of pairs of nodes with positive mutual connection probability $p_{i_0j_0}^{\leftrightarrow}$ (contributing to an increase of the attainable value for $\langle r \rangle_{\min}$).
In our infinite-mean fitness case, typical values of the fitness are spread so broadly over nodes that they effectively hinder the emergence of antireciprocal patterns in the network.

\begin{figure}[t!]
    \centering
    \includegraphics[width=0.4\textwidth]{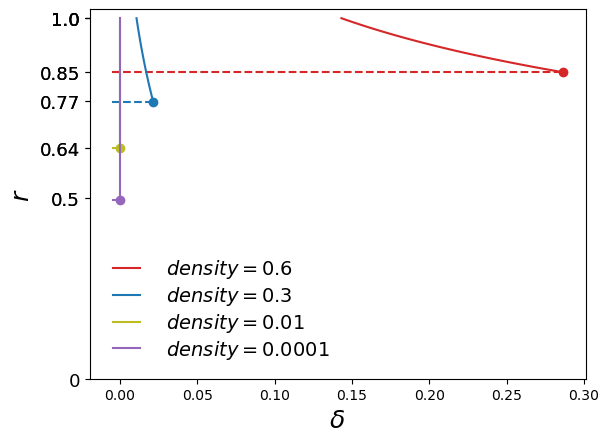}%
    \caption{\small\textbf{Reciprocity $r$ in the annealed case as a function of the parameter $\delta$.}
    Solid lines illustrate, for four different values of link density ($\overline{\kappa} = 0.0001, 0.01, 0.3, 0.6 $, corresponding to $\epsilon=2.83\,e^{-10}, 1.83\,e^{-06},0.01, 0.14$) the behaviour of the function $\langle r \rangle$ in~\eqref{eq: reciprocity}, while the full circles indicate the values achieved in the areciprocal case $\delta_{\text{rand}} = 2 \epsilon$ (where  $\langle r \rangle_{\text{rand}}$ is given by~\eqref{eq: rec_rand}). The dashed lines indicate the antireciprocal limit $\langle r \rangle_{\min}=\langle r \rangle_{\delta_{\max}}$, with $\delta_{\max} = \min_{i,j}{ \left(-\frac{\ln(2 e^{-\epsilon x_i x_j}-1)}{x_i x_j}\right)}$. 
    In each plot, $\delta$ varies in the range of its domain of existence $\delta \in \left[\epsilon, \delta_{\max} \right]$, which is dependent upon the specific realization of fitness values (which is fixed in all plots and amounting to $5000$ values) and, clearly, on the parameter $\epsilon$ (whose value is found so to enforce the desired density of links $\overline{\kappa}$).
    }
    \label{fig: annealed reciprocity}
\end{figure}

In analogy with our analysis in Sec.~\ref{section: ER}, we can evaluate the expected reciprocity $\langle r \rangle(\epsilon, \delta)$ in Eq.~\eqref{eq: reciprocity} for L\'{e}vy-distributed fitness values as a function of the parameter $\delta$ in the domain established by the conditions in Eq.~\eqref{eq: CE}, for different values of the parameter $\epsilon$ (i.e. for a varying density of links).
In Fig.~\ref{fig: annealed reciprocity} the outcome of this procedure is illustrated for four different values of the expected link density $\overline{\kappa} = \frac{1}{N(N-1)}\sum_{i,j\neq i} p_{ij}$.
As the figure shows, maximal reciprocity ($\langle r \rangle=1$) can always be achieved (specifically, when  $\delta = \delta_{\min} = \epsilon$, yielding $p_{ij}^{\rightarrow} = 0$ and $p_{ij} = q_{ij} \; \forall\, i, j$).
By contrast, $\langle r\rangle$  can never reach antireciprocated values compatibly with the conditions in Eq.~\eqref{eq: CE}: $\delta_{\max}$ is indeed too small to allow for global values of reciprocity below the areciprocal benchmark, due to the coexistence of nodes with very low fitness (which yield $\delta_{ij}^{\max}\approx 2 \epsilon$) and nodes with very large fitness (which can establish unreciprocated links only for very large $\delta$).

These results show that the annealed random networks that remain invariant under aggregations have non-negative reciprocity: they can only be areciprocal or reciprocal. In a certain sense, antireciprocity does not `resist' aggregation in the annealed setting.

\begin{figure*}[t!]
    \centering
    \includegraphics[width = 0.8\textwidth]{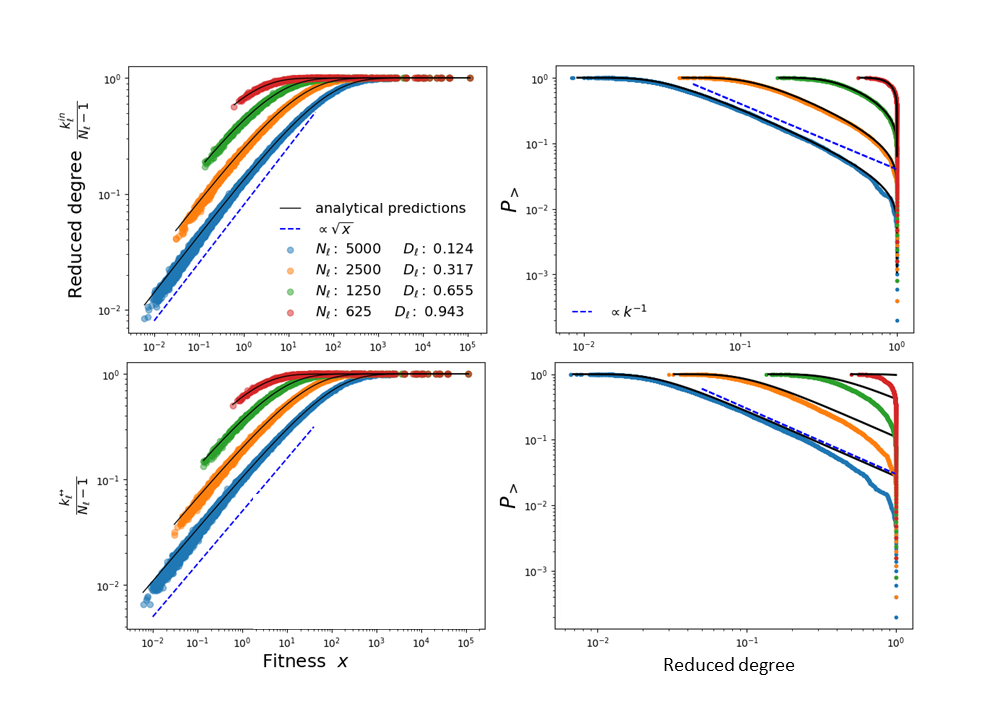} %
    \caption{\small\textbf{Degrees and reciprocal degrees in the annealed DSIM}. We show the outcome of a simulation obtained from one realization of a synthetic network of $N_0=5000$ nodes at four different levels of aggregation ($\ell = 0, 1, 2, 3$). This is performed by drawing the network realization from the graph probability in~\eqref{Deq_product2} and then progressively coarse-graining it via the renormalization rules in~\eqref{Deq:cg rule}, after having fixed the hierarchy of partitions in the most homogeneous way: at each step $\ell$, $N_{\ell}$ blocks are created by randomly grouping nodes in pairs so that we progressively halve the number of nodes  $N_\ell = N_0 2^{-\ell}$.
    In the two figures on the left,  the reduced in-degree $\kappa^{\text{in}}_\ell$ (top) and reduced reciprocal degree $\kappa^{\leftrightarrow}_\ell$ (bottom) are shown as functions of the fitness values and compared to the analytical predictions in~\eqref{eq: kappa_in(x)} and~\eqref{eq: k_rec VS x}. In the two figures on the right, the cumulative distributions of the reduced in-degree $\kappa^{\text{in}}$ and reduced reciprocal degree $k^{\leftrightarrow}$ are compared with the analytical expectations shown in equations~\eqref{eq: cumulative in-deg} and~\eqref{eq: cumulative rec-deg}. As a reference, the dashed lines in the right panels are proportional to $\kappa^{-1}$, corresponding to $\kappa^{-2}$ for the non-cumulative pdf.}
    \label{fig: degrees}
\end{figure*}

\subsection{Expected topological properties}
We now calculate the expected degrees (in-, out-, and reciprocated degrees) and their distributions.
Note that the expected in-degree $k_{i_\ell}^{\text{in}}$, in this symmetrical setting, is equivalent to the expected out-degree $k_{i_\ell}^{\text{out}}$.
In what follows, we will consider for convenience the rescaled quantities $\kappa_\ell^{in}(x) \equiv \frac{k_\ell^{in}(x)}{N_\ell-1}= \frac{k_\ell^{out}(x)}{N_\ell-1}$ and  $\kappa_\ell^{\leftrightarrow}(x) \equiv \frac{k_\ell^{\leftrightarrow}(x)}{N_\ell-1}$.

As shown in SI, the expected rescaled in-degree of a node with fitness $x$ at level $\ell$ is given by
\begin{equation}\label{eq: kappa_in(x)}
    \kappa_\ell^{in}(x) =  1- e^{-\sqrt{2 \epsilon \gamma_\ell x}},
\end{equation}
a relationship confirmed via numerical simulations in the top left panel of Fig.~\ref{fig: degrees}.
Inverting the above equation allows to calculate the resulting expected in-degree distribution exactly (see SI) as
\begin{equation} \label{eq: cumulative in-deg}
    P_\ell(\kappa^{in}) = \frac{2}{\sqrt{\pi}} \frac{\sqrt{\epsilon} \gamma_{\ell}}{1-\kappa^{in}} \frac{e^{- \frac{\epsilon \gamma_{\ell}^2}{\log^2(1-\kappa^{in})}}}{\log^2(1-\kappa^{in})}.
\end{equation}
Note that this distribution has a power-law regime proportional to $\kappa^{-2}$, followed by a density-dependent cut-off.
The top right panel of Fig.~\ref{fig: degrees} illustrates the good agreement between numerical simulations and analytical predictions at four different levels of resolution.

Along the same line, the dependence of the expected rescaled reciprocal degree $\kappa_{\ell}^{\leftrightarrow} \equiv \frac{k_{\ell}^{\leftrightarrow}}{N_\ell-1}$ on the fitness can be calculated exactly:
\begin{equation} \label{eq: k_rec VS x}
\kappa_{\ell}^{\leftrightarrow}(x) =  1 - 2 e^{-\sqrt{2 \epsilon \gamma_{\ell} x}} + e^{-\sqrt{2 \delta  \gamma_{\ell} x}}.
\end{equation}
This is confirmed in the bottom left panel of Fig.~\ref{fig: degrees}.
In this case, however, the expression cannot be inverted unless a specific relation is enforced between the parameters  $\epsilon$ and $\delta$, i.e.  $\delta = 4 n \epsilon$, with $n = 1, 2, \dots$ (this relation might in general be incompatible with the conditions in Eq.~\eqref{eq: CE}). 
Nevertheless, we can estimate the cumulative distribution of rescaled reciprocated degrees by considering the behaviour of  $\kappa_{\ell}^{\leftrightarrow}(x)$ for small values of $x$, i.e. $\kappa_{\ell}^{\leftrightarrow}(x) \sim \sqrt{2 \gamma_{\ell} x} \;(2 \sqrt{\epsilon} - \sqrt{\delta})$. 
In this regime the inverse function $x(k_{\ell}^{\leftrightarrow})$ can easily be found, and the resulting (approximate) reciprocal degree distribution reads
\begin{equation} \label{eq: cumulative rec-deg}
\Tilde{P}_\ell(k^{\leftrightarrow}) \approx \frac{2}{\sqrt{\pi}}   \frac{\gamma_\ell \left(2\sqrt{\epsilon} - \sqrt{\delta}\right)}{\left(k^{\leftrightarrow}\right)^2}  e^{-\left( 
    \frac{\gamma_{\ell}(2 \sqrt{\epsilon} - \sqrt{\delta})}{k^{\leftrightarrow}}\right)^2} .
\end{equation}
As confirmed in the bottom right panel of Fig.~\ref{fig: degrees}, this approximate result reproduces the bulk of the distribution, while failing in the tails as expected (the approximation being valid only for small values of $x$).

\begin{figure}[t!]
    \centering
    \includegraphics[width=0.4\textwidth]{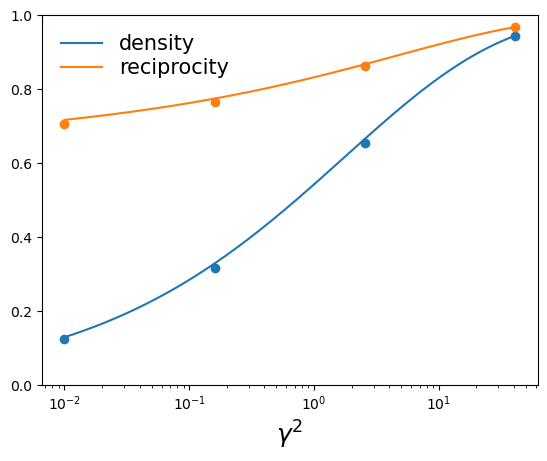}
    \caption{\small\textbf{Link density and reciprocity along the coarse-graining flow}. The blue and the orange lines correspond to the analytical predictions, respectively,  of the link density $\langle \bar{\kappa}_\ell \rangle$ and reciprocity $\langle r_\ell \rangle$ in equations~\eqref{eq:Meijers_k}-\eqref{eq:Meijers_r}. The full circles correspond to the realized values of $\langle \bar{\kappa}_\ell \rangle$ and  $\langle r_\ell \rangle$ in a simulation where a synthetic network of $N_0=5000$ nodes is generated at level $\ell = 0$ and then progressively coarse-grained.
    At each iteration, nodes are randomly merged in pairs ($b_\ell=2$) so that the parameter $\gamma_\ell$ from level $0$ to $\ell +1$ scales as $\gamma_{\ell +1} = 2^{2 \ell} \gamma_{0}$. Parameters $\eta$ and $\delta$ are fixed, respectively, to $0.1$ and $\delta_{max} =0.20000037$ (the latter yielding the minimally reciprocated networks, given $\epsilon = 0.1$ and $\{ x_{i_0}\}$).}
    \label{fig:Meijers}
\end{figure}

Finally, we look at the expected density of links $\langle \overline{\kappa}_{\ell} \rangle$ (excluding self-loops) and the expected reciprocity $\langle r \rangle$. 
An expression for the density is found from
\begin{equation} \label{Seq: almostD}
\begin{split}
    \langle \bar{\kappa}_\ell \rangle &= \int_0^{\infty} \mathrm{d}x \int_0^{\infty} \mathrm{d}y \,f(x,y) \varphi (x|\gamma_\ell) \varphi (y|\gamma_\ell)\\
   &= 1-\int_0^{\infty}  e^{-\sqrt{2 \epsilon \gamma_\ell x}} \varphi (x|\gamma_\ell) \mathrm{d}x,\\
\end{split}
\end{equation}
which can be written in terms of the Meijer-$G$ function: 
\begin{equation} \label{eq:Meijers_k}
\langle \bar{\kappa}_\ell\rangle = 1-\frac{\sqrt{\eta}\gamma_\ell}{2\pi}\MeijerG{3}{0}{0}{3}{\cdot}{-1/2,0,0}{ {\frac{\epsilon \gamma_\ell^2}{4}}}.
\end{equation}
Analogously, we can express the expected reciprocity as
\begin{eqnarray} 
\langle r_\ell \rangle &=&  \frac{\int_0^{\infty} \int_0^{\infty} f^{\leftrightarrow}(x,y) \varphi(x|\gamma_\ell) \varphi(y|\gamma_\ell) \mathrm{d}x \mathrm{d}y}{\langle \bar{\kappa}_\ell\rangle}\nonumber \\ 
&=&2 - \frac{1-\frac{\sqrt{\delta}\gamma_\ell}{2\pi}\MeijerG{3}{0}{0}{3}{\cdot}{-1/2,0,0}{ {\frac{\delta \gamma_\ell^2}{4}}} }{\langle \bar{\kappa}_\ell\rangle}.\label{eq:Meijers_r}
\end{eqnarray}
We display the above analytical results in Fig.~\ref{fig:Meijers}, where the functional dependence, shown in Eqs.~\eqref{eq:Meijers_k} and \eqref{eq:Meijers_r}, of $\langle \overline{\kappa}_\ell \rangle$ and $\langle r_\ell \rangle$  on the scale parameter $\gamma_\ell$ is illustrated, along with the respective values obtained at four different levels of aggregation of a synthetic network drawn from the graph probability in~\eqref{Deq_product2}  and then progressively coarse-grained by merging pairs of randomly chosen nodes ($b_\ell=2$). 
Again, numerical simulations agree perfectly with the analytical calculations.

\section{Conclusions}
Motivated by the problem of multiscale modelling, a recent work proposed a Scale-Invariant random graph model (SIM) built on the principles of renormalization group theory as defined in the space of configurations~\cite{Elenulla}. 
We have shown here that such a Scale-Invariant Model can be extended coherently to directed networks, leading to the Directed Scale-Invariant Model (DSIM).
Importantly, besides being scale-invariant, the DSIM can account for nontrivial reciprocity.

As in the undirected formulation, the model can generate any directed network in two equivalent ways: either hierarchically (by generating a finer-grained network and then coarse-graining it through a sequence of nested partitions) or directly (using appropriately renormalized parameters). 
In the directed formulation, these parameters include three global parameters (tuning the overall density of directed links, reciprocated links and self-loops), four vectors of fitness variables (separately determining the tendency of nodes to establish incoming,  outgoing, reciprocated and self-connections) and, if useful, a matrix of dyadic factors representing distances, similarities, communities, etc. 
A sufficient condition to ensure that the connection probabilities are indeed scale-invariant even in the presence of dyadic interactions is to consider \emph{metaultrametric} distances compatible with the hierarchy of partitions defining the coarse-graining procedure.

After  deriving the renormalization rules for networks with arbitrary reciprocity, we have considered various examples, including the first directed multiscale model of the international trade network and an annealed model, where scale-free networks with positive reciprocity emerge spontaneously from coarse-graining.

Applying the model to the directed ITN demonstrated its consistency with the undirected formalism (accurately reproducing topological features of the undirected projection of the ITN), while at the same time capturing nontrivial directed topological and spectral properties related to strong reciprocity patterns across hierarchical aggregation levels.
To our knowledge, this application of our method represents the first multiscale model replicating the structural and spectral properties of the empirical directed ITN. 
The model is very parsimonious as it requires fitting only 2 free global parameters to replicate several local structural properties, as well as the elliptical shape of the spectral distribution, across arbitrary coarse-grainings of the network.

In the annealed scenario, we focused on a specific case, where the connection probabilities are symmetric (in that the different fitness parameters assigned to each node are taken to be identical) and the fitness values are drawn from a L\'{e}vy distribution.  
The high heterogeneity of this distribution suppresses the antireciprocal regime, suggesting that annealed scale-invariant random graphs are intrinsically reciprocal.
In such a setting, we have shown that in-degree, out-degree and reciprocal degree distributions feature a power-law decay $\propto k^{-2} $ followed by a density-dependent cutoff.
Therefore, the sole requirement of scale-invariance automatically produces directed scale-free graphs with positive reciprocity.

These results show that it is possible to define, without any notion of geometric embedding, a consistent aggregation-invariant description of directed networks. Future research should address further generalizations, such as higher-dimensional quenched fitness values and multivariate $\alpha$-stable distributions for the different annealed fitness types.

\begin{acknowledgments}
This work is supported by the European Union - NextGenerationEU - National Recovery and Resilience Plan (Piano Nazionale di Ripresa e Resilienza, PNRR), project `SoBigData.it - Strengthening the Italian RI for Social Mining and Big Data Analytics' - Grant IR0000013 (n. 3264, 28/12/2021). 
This work is also supported by the project NetRes - `Network analysis of economic and financial resilience', Italian DM n. 289, 25-03-2021 (PRO3 Scuole), CUP D67G22000130001 (\url{https://netres.imtlucca.it}).
DG acknowledges support from the Dutch Econophysics Foundation (Stichting Econophysics, Leiden, the Netherlands) and the Netherlands Organization for Scientific Research (NWO/OCW).
\end{acknowledgments}

\bibliography{reference2}


%
%
\clearpage
\newpage
\setcounter{equation}{0}

\setcounter{figure}{0}
\setcounter{table}{0}
\setcounter{page}{1}
\setcounter{section}{0}

\renewcommand{\theequation}{S\arabic{equation}}
\renewcommand{\thetable}{S\arabic{table}}
\renewcommand{\thefigure}{S\arabic{figure}}
\renewcommand{\thesection}{S.\Roman{section}} 
\renewcommand{\thesubsection}{\thesection.\roman{subsection}}

\onecolumngrid
{\center
\textbf{SUPPLEMENTARY INFORMATION}\\
$\quad$\\
accompanying the paper\\
\emph{``Geometry-free renormalization of directed networks: scale-invariance and reciprocity''}\\
by M. Lalli and D. Garlaschelli\\
$\quad$\\
$\quad$\\
}

\section{Graph probability and partition function}\label{SI-section:P and Z}
Here we show the calculation of the full graph probability in the Directed Scale Invariant Model (DSIM). Given the quantities defined in the main text, we proceed as follows:
\begin{eqnarray}
P\big(\mathbf{A}^{(\ell)} \big) &=& \prod_{i_{\ell}=1}^{N_{\ell}}  \prod_{j_{\ell}=1}^{i_{\ell}} \left(p_{i_{\ell}j_{\ell}}^{\rightarrow} \right)^{{a_{i_{\ell}j_{\ell}}^{\rightarrow}} } \left(p_{i_{\ell}j_{\ell}}^{\leftarrow}  \right)^{a_{i_{\ell}j_{\ell}}^{\leftarrow}}\left(p_{i_{\ell}j_{\ell}}^{\leftrightarrow}  \right)^{a_{i_{\ell}j_{\ell}}^{\leftrightarrow}}\left(p_{i_{\ell}j_{\ell}}^{\nleftrightarrow} \right)^ {a_{i_{\ell}j_{\ell}}^{\nleftrightarrow} }\nonumber\\
 &=& \prod_{i_\ell =1}^{N_\ell}   \prod_{j_\ell =1}^{i_\ell}  \left( \frac{p^{\rightarrow}_{i_\ell j_\ell}}{p^{\nleftrightarrow}_{i_\ell j_\ell}}\right)^{a^{\rightarrow}_{i_\ell j_\ell}} \left( \frac{p^{\leftarrow}_{i_\ell j_\ell}}{p^{\nleftrightarrow}_{i_\ell j_\ell}}\right)^{a^{\leftarrow}_{i_\ell j_\ell}}  \left( \frac{p^{\leftrightarrow}_{i_\ell j_\ell}}{p^{\nleftrightarrow}_{i_\ell j_\ell}}\right)^{a^{\leftrightarrow}_{i_\ell j_\ell}} p^{\nleftrightarrow}_{i_\ell j_\ell} \nonumber\\
&=& Q\prod_{i_\ell =1}^{N_\ell}   \prod_{j_\ell =1}^{i_\ell}  \left( \frac{p^{\rightarrow}_{i_\ell j_\ell}}{p^{\nleftrightarrow}_{i_\ell j_\ell}}\right)^{a^{\rightarrow}_{i_\ell j_\ell}} \left( \frac{p^{\leftarrow}_{i_\ell j_\ell}}{p^{\nleftrightarrow}_{i_\ell j_\ell}}\right)^{a^{\leftarrow}_{i_\ell j_\ell}}  \left( \frac{p^{\leftrightarrow}_{i_\ell j_\ell}}{p^{\nleftrightarrow}_{i_\ell j_\ell}}\right)^{a^{\leftrightarrow}_{i_\ell j_\ell}},\label{Seq:piccetta} 
\end{eqnarray}
where  $Q$ is a constant quantity defined as 
\begin{eqnarray}
    Q    &\equiv &\prod_{i_\ell =1}^{N_\ell}   \prod_{j_\ell =1}^{i_\ell}  p^{\nleftrightarrow}_{i_\ell j_\ell}\nonumber\\
&=&    \prod_{i_{\ell}=1}^{N_{\ell}} e^{-\frac{\delta}{2} x_{i_{\ell}}^2 f(d_{i_{\ell}i_{\ell}}) - \eta w_{i_{\ell}}}  \prod_{j_{\ell} = 1}^{i_{\ell}-1} e^{-\delta x_{i_{\ell}} x_{j_{\ell}} f(d_{i_{\ell}j_{\ell}})} \nonumber\\
    &=&  e^{-\frac{\delta}{2}  \sum_{i_\ell =1}^{N_\ell} \sum_{j_\ell =1}^{N_\ell} x_{i_{\ell}} x_{j_{\ell}} f(d_{i_{\ell}j_{\ell}}) - \eta \sum_{i_\ell =1}^{N_\ell} w_{i_\ell} } \nonumber \\
    &=& e^{-\frac{\delta}{2}  x_{i_{\infty}}^2 f(d_{i_{\infty},i_{\infty}}) - \eta  w_{i_{\infty}} } \nonumber\\
    &=& 1-p_{i_{\infty}i_{\infty}},\nonumber
\end{eqnarray}
with $x_{i_\infty},\, w_{i_{\infty}}$ and $f(d_{i_\infty,i_\infty})$ being scale-invariant quantities, formally defined as the total `undirected' fitness, the total `self-loop' fitness and  the weighted average of the dyadic factor among all pairs of nodes,  respectively:
\begin{eqnarray}
x_{i_\infty}&\equiv&\sum_{i_\ell=1}^{N_\ell}x_{i_\ell}=\sum_{i_0=1}^{N_0}x_{i_0},\label{xinfty}\\
w_{i_{\infty}} &\equiv& \sum_{i_{\ell}=1}^{N_{\ell}} w_{i_{\ell}}=\sum_{i_0=1}^{N_0}w_{i_0} ,\label{winfty}\\
f(d_{i_\infty,i_\infty})&\equiv&\sum_{i_{\ell}=1}^{N_\ell}\sum_{j_{\ell}=1}^{N_\ell} \frac{x_{i_{\ell}} x_{j_{\ell}} f\big(d_{i_{\ell},j_{\ell}}\big)}{x^2_{i_\infty}}=\sum_{i_{0}=1}^{N_0}\sum_{j_{0}=1}^{N_0} \frac{x_{i_{0}} x_{j_{0}} f\big(d_{i_{0},j_{0}}\big)}{x^2_{i_\infty}}\label{finfty}.
\end{eqnarray}
In other words, $i_\infty$ formally represents the single node produced at $\ell = \infty$ by the `ultimate' partition that places all nodes in a single block. Node $i_{\infty}$ is connected to itself unless the original network is completely devoid of links and self-loops. Clearly, recalculating $Q$ starting from any other hierarchical level $\ell'\ne \ell$ would return the same value, which proves its scale invariance.

In a (formal) analogy with Exponential Random Graphs~\cite{wasserman1994social,park2004statistical,cimini2019statistical}, the graph probability $P\left(\mathbf{A}^{(\ell)}  \right) $ can be rewritten in terms of an effective Hamiltonian $\mathcal{H}^{(\ell)}_\textrm{eff}$ and a partition function $\mathcal{Z}^{(\ell)}$, in a way that highlights a connection with real-space (Kadanoff) renormalization \cite{kadanoff2000statistical,wilson1983renormalization}:
\begin{equation}\label{Seq:erg}
P\left(\mathbf{A}^{(\ell)}  \right) = \frac{e^{-\mathcal{H}^{(\ell)}_\textrm{eff}\left(\mathbf{A}^{(\ell)} \right)}}{\mathcal{Z}^{(\ell)}},
\end{equation}
where 
\begin{equation}\label{eq:Heff}
\mathcal{H}^{(\ell)}_\textrm{eff}\left(\mathbf{A}^{(\ell)}  \right) = - \sum_{i_\ell =1}^{N_\ell}    \sum_{j_\ell =1}^{i_\ell}  \left[ a^{\rightarrow}_{i_\ell j_\ell} \log{ \left( \frac{p^{\rightarrow}_{i_\ell j_\ell}}{p^{\nleftrightarrow}_{i_\ell j_\ell}}\right)} +   a^{\leftarrow}_{i_\ell j_\ell} \log{\left( \frac{p^{\leftarrow}_{i_\ell j_\ell}}{p^{\nleftrightarrow}_{i_\ell j_\ell}}\right)} + a^{\leftrightarrow}_{i_\ell j_\ell} \log{\left( \frac{p^{\leftrightarrow}_{i_\ell j_\ell}}{p^{\nleftrightarrow}_{i_\ell j_\ell}}\right)} \right],
\end{equation}
and
\begin{eqnarray}
\mathcal{Z}^{(\ell)}  &\equiv&\sum_{\{\mathbf{A}^{(\ell)}\}} e^{-\mathcal{H}^{(\ell)}_\textrm{eff}\left(\mathbf{A}^{(\ell)} \right)} \\
& =&\sum_{\{\mathbf{A}^{(\ell)}\}} \prod_{i_\ell =1}^{N_\ell}    \prod_{j_\ell =1}^{i_\ell} \left( \frac{p^{\rightarrow}_{i_\ell j_\ell}}{p^{\nleftrightarrow}_{i_\ell j_\ell}}\right)^{a^{\rightarrow}_{i_\ell j_\ell}} \left( \frac{p^{\leftarrow}_{i_\ell j_\ell}}{p^{\nleftrightarrow}_{i_\ell j_\ell}}\right)^{a^{\leftarrow}_{i_\ell j_\ell}}  \left( \frac{p^{\leftrightarrow}_{i_\ell j_\ell}}{p^{\nleftrightarrow}_{i_\ell j_\ell}}\right)^{a^{\leftrightarrow}_{i_\ell j_\ell}} \\
& =&\prod_{i_\ell =1}^{N_\ell}    \prod_{j_\ell =1}^{i_\ell} \sum_{   \{ (a^{\rightarrow}_{i_\ell j_\ell}, a^{\leftarrow}_{i_\ell j_\ell},a^{\leftrightarrow}_{i_\ell j_\ell} )\} }   \left( \frac{p^{\rightarrow}_{i_\ell j_\ell}}{p^{\nleftrightarrow}_{i_\ell j_\ell}}\right)^{a^{\rightarrow}_{i_\ell j_\ell}} \left( \frac{p^{\leftarrow}_{i_\ell j_\ell}}{p^{\nleftrightarrow}_{i_\ell j_\ell}}\right)^{a^{\leftarrow}_{i_\ell j_\ell}}  \left( \frac{p^{\leftrightarrow}_{i_\ell j_\ell}}{p^{\nleftrightarrow}_{i_\ell j_\ell}}\right)^{a^{\leftrightarrow}_{i_\ell j_\ell}} \label{qui}\\
& =&\prod_{i_\ell =1}^{N_\ell}    \prod_{j_\ell =1}^{i_\ell}   \left( \frac{p^{\rightarrow}_{i_\ell j_\ell}}{p^{\nleftrightarrow}_{i_\ell j_\ell}} + \frac{p^{\leftarrow}_{i_\ell j_\ell}}{p^{\nleftrightarrow}_{i_\ell j_\ell}}  +    \frac{p^{\leftrightarrow}_{i_\ell j_\ell}}{p^{\nleftrightarrow}_{i_\ell j_\ell}}+ 1 \right) \label{qua}\\
& =&\prod_{i_\ell =1}^{N_\ell}    \prod_{j_\ell =1}^{i_\ell}   \frac{1 }{p^{\nleftrightarrow}_{i_\ell j_\ell}}  \\
& =& Q^{-1}.
\end{eqnarray}
In going from~\eqref{qui} to~\eqref{qua}, we have used $\{ (a^{\rightarrow}_{i_\ell j_\ell}, a^{\leftarrow}_{i_\ell j_\ell},a^{\leftrightarrow}_{i_\ell j_\ell} )\}=(1,0,0),(0,1,0),(0,0,1),(0,0,0)$ following from the mutual exclusiveness of the different dyads involving the same two nodes.
Being equal to $Q^{-1}$, which is scale-invariant, the partition function is exactly scale-invariant itself, in line with Kadanoff's construction in real-space renormalization~\cite{kadanoff2000statistical}.\\ \\

As explained in the main text, to ensure that the graph  probability in equation~\eqref{Seq:piccetta} is well defined at any level of aggregation, the following inequalities must hold $\forall\; \ell$:
\begin{equation} \label{Seq: CE}
    \delta^{(i_\ell,j_\ell)}_\textrm{min} \leq \delta  \leq \delta^{(i_\ell,j_\ell)}_\textrm{max}   \quad \forall i_\ell \neq j_\ell
\end{equation}
where
\begin{equation*}
    \delta^{(i_\ell,j_\ell)}_\textrm{max} \equiv \begin{cases}
        - \frac{\ln{ \left( e^{-\epsilon y_{i_\ell} z_{j_\ell} f(d_{i_\ell j_\ell})} + e^{-\epsilon y_{j_\ell} z_{i_\ell} f(d_{i_\ell j_\ell})} - 1 \right)}}{x_{i_\ell} x_{j_\ell} f(d_{i_\ell j_\ell}) }  & \text{ if } \; 1\!-\!p_{i_\ell j_\ell}\!-\!p_{j_\ell i_\ell} > 0\\
        +\infty & \text{ if } \; 1\!-\!p_{i_\ell j_\ell}\!-\!p_{j_\ell i_\ell}  \leq 0
    \end{cases}
\end{equation*}
while
\begin{equation*}
    \delta^{(i_\ell,j_\ell)}_\textrm{min} \equiv\frac{ \epsilon \max{(y_{i_\ell} z_{j_\ell}, y_{j_\ell} z_{i_\ell})}}{x_{i_\ell} x_{j_\ell}}.
\end{equation*}
The above condition must hold for all pairs $i_\ell$, $j_\ell$ simultaneously, which also means for all hierarchical levels $\ell\ge 0$. 
Note that a sufficient condition for the above inequalities to hold at any hierarchical level is that $\delta^{(i_\ell,j_\ell)}_\textrm{max} $ is non-decreasing as $\ell$ increases, while $\delta^{(i_\ell,j_\ell)}_\textrm{min} $ is a decreasing function, implying:
\begin{eqnarray}
    \!\!\!\!\!\!&\frac{ \;\max(y_{i_{\ell \!+\!1}} z_{j_{\ell \!+ \!1}},y_{i_{\ell \!+ \!1}} z_{j_{\ell \!+ \!1}} )}{x_{i_{\ell \!+ \!1}} x_{j_{\ell \!+ \!1}}} \leq \frac{ \;\max(y_{i_{\ell}} z_{j_{\ell}},y_{i_{\ell}} z_{j_{\ell}} )}{x_{i_{\ell}} x_{j_{\ell}}},&\!\!\!\!\!\! \label{Deq: CE coarse-grained}\\
    \!\!\!\!\!\!&\frac{{ \left[ e^{-\epsilon \, y_{i_{\ell \!+ \!1}} z_{j_{\ell \!+ \!1}}f(d_{i_{\ell+1}j_{\ell+1}})} + e^{-\epsilon \, y_{j_{\ell \!+ \!1}} z_{i_{\ell \!+ \!1}}f(d_{i_{\ell+1}j_{\ell+1}})} \!-\!1 \right]^{x_{i_{\ell}} x_{j_{\ell}}f(d_{i_{\ell}j_{\ell}})}}} 
    { \left[ e^{-\epsilon \, y_{i_{\ell}} z_{j_{\ell}}f(d_{i_{\ell}j_{\ell}})} + e^{-\epsilon \, y_{j_{\ell}} z_{i_{\ell}}f(d_{i_{\ell}j_{\ell}})} \!-\!1 \right]^{x_{i_{\ell \!+ \!1}} x_{j_{\ell \!+ \!1}}f(d_{i_{\ell+1}j_{\ell+1}})}} 
    \leq 1.&\!\!\!\!\!\!
\end{eqnarray} 


\section{GDP, ultrametric distances and trade data} \label{SI: data}
Here we specify the procedure used to gather trade, GDP and distance data in our analysis of the International Trade Network (ITN). 

For defining the fitness values and the empirical network we relied on the expanded trade dataset developed by K. S. Gleditsch~\cite{gleditsch2002expanded}, which provides estimates of the bilateral trade flows between independent states between the years from 1948 to 2000, together with estimates of their Gross Domestic Products (GDPs) per capita and population. 
The GDP estimates are originally based on figures from Penn World Tables~\cite{PWT}, produced by the Center for International Comparisons at the University of Pennsylvania, and from reports of Central Intelligence Agency (CIA)~\cite{CIA}.
The outcome consists of 185 countries, whose \textit{real GDPs per capita} are given for all years in constant US dollars (with base 1996), along with population figures in units of 1000s.
We obtained the real \emph{total GDP} value of each country by multiplying its GDP value per capita by its population. 
Such quantities are naturally additive (i.e., scale linearly with the population) and are thus suitable to serve as nodes' fitness.

For defining the binary asymmetric matrix $\Tilde{\mathbf{A}}^{(0)}$ representing our empirical ITN at level $\ell=0$, we used the final estimates of the international trade flows provided in Gleditsch database, obtained from a refinement of the Direction of Trade (DOT) of the International Monetary Fund (IMF) data~\cite{IMF} where missing data and suspicious zero had been managed with figures from the World Export Data (WED)~\cite{WED} or through ad hoc procedures (for which we refer to the original reference~\cite{gleditsch2002expanded}). 
As described in the main text, if a positive trade is estimated in the considered year (2000) from country $i_0$ to the country $j_0$ then we posit $a_{i_0j_0} = 1$, otherwise $a_{i_0j_0} = 0$.

As explained in the main text, the pairwise distances between countries are set as the subdominant metaultrametric distances of certain population-averaged geographical distances provided within the BACI-CEPII GeoDist database~\cite{mayer2011notes} (which, in its turn, relies on the World Gazetteer  website~\cite{populationgeodist} for population data). These are bilateral inter-country distances measured as the population-based averages among the most populated pairs of cities across each pair of countries. The general formula for calculating this effective distance $d_{i_0,j_0}$ between countries $i_0$ and $j_0$, developed by Head and Mayer~\cite{head2002illusory}, is given by:

\begin{equation} \label{Seq: Meyer dist}
d_{i_0,j_0}=\left(\frac{\sum_{k\in i_0}\sum_{l\in j_0}\mathrm{POP}_k\mathrm{POP}_l\,d^{\,-1}_{k.l}}{\sum_{k\in i_0}\sum_{l\in j_0}\mathrm{POP}_k\mathrm{POP}_l}\right)^{-1}
\end{equation}
where the sum runs on pairs of the most populated internal agglomerations (cities, towns and places) across $i_0$ and $j_0$, while $\mathrm{POP}_k$ denotes the demographic population of agglomeration $k$.

\begin{figure}
    \centering
    \includegraphics[width = 0.9\textwidth]{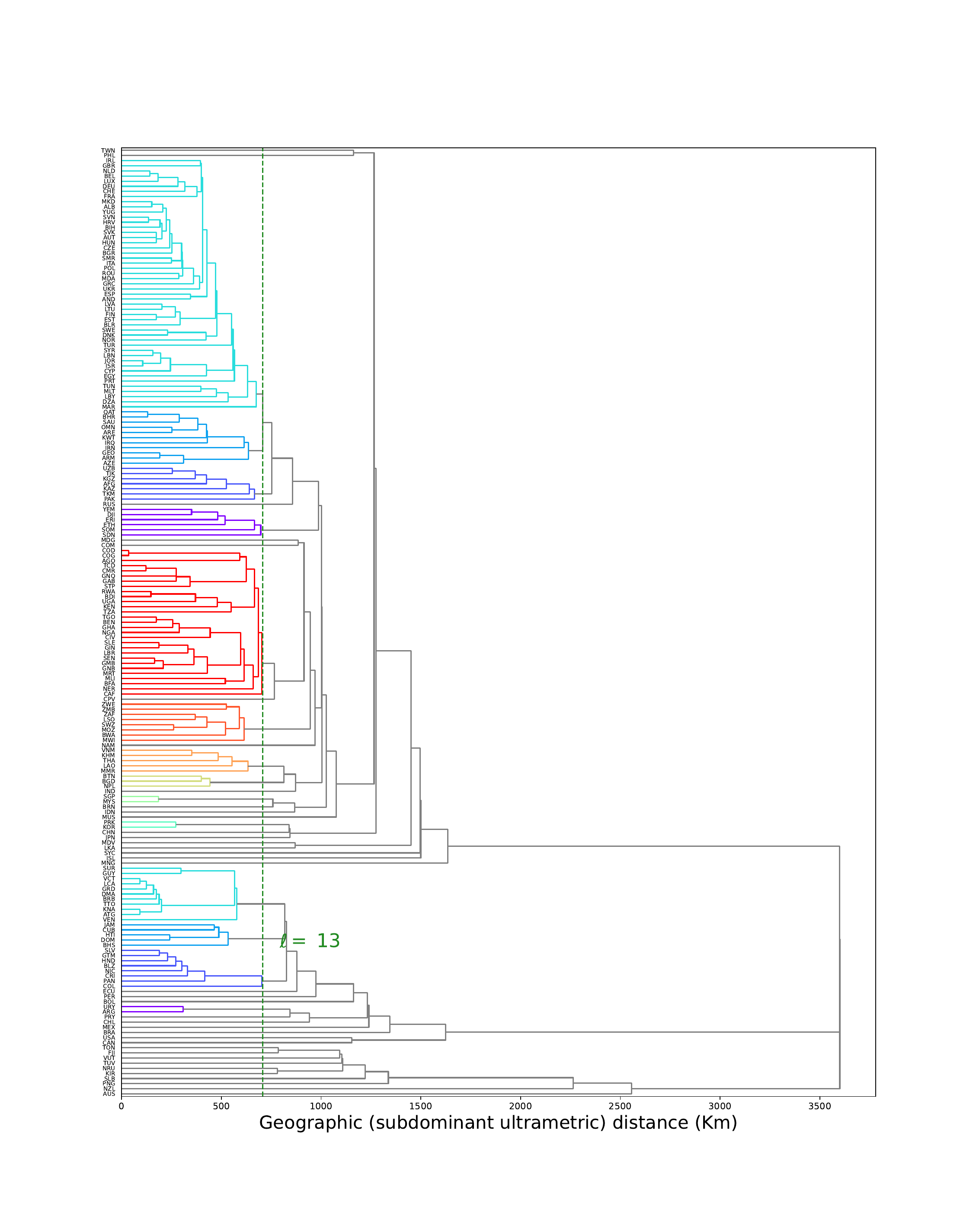}
    \caption{\small \textbf{Dendrogram of world countries from their geographical distances using single-linkage hierarchical clustering.} The dendrogram can be used to produce any desired sequence $\{\mathbf{\Omega}_\ell\}_{\ell\ge 0}$ of geographically nested partitions, via either single-scale (here, unique vertical lines) or multi-scale (multiple vertical lines, but monophyletic) `cuts'. In our analysis we considered 17 single-scale cuts at various metaultrametric distances $\{h_\ell\}_{\ell=0}^{16}$ (with $h_0=0$) producing a hierarchy $\{\mathbf{\Omega}_\ell\}_{\ell=0}^{16}$ of 17 partitions and a corresponding sequence of block-countries with $N_0=185$ and $N_\ell=180-10\ell$ for $\ell=1,16$. For instance, a cut at level $\ell=13$ (dashed line) yields 50 block-countries that correspond to the 50 branches drawn in different colors.}
    \label{figSI: dendro}
\end{figure}

As explained in the main text, the dyadic parameters considered in the model are required to be ultrametric (or, more in general, metaultrametric) allowing us to define the coarse-graining flow of the empirical network by progressively cutting the associated dendrogram  at always larger heights while keeping invariant such parameter.
To minimize the distortion between the original inter-country distance $d_{i_0,j_0}$ and its metaultrametric approximation $d^{\text{mu}}_{i_0,j_0}$,  we consider the subdominant ultrametric distance, i.e., the largest among all ultrametric that are less than or equal to the original metric. 
These can be obtained via a single-linkage hierarchical clustering performed by using the original distances $d_{i_0 j_0}$ as measure of dissimilarity. 
Hierarchical clustering algorithms start from the partition of the dataset into singleton nodes and merge step by step the current pair of mutually closest nodes into a new node until there is one final node left. 
Several clustering schemes share this procedure as a common definition, but differ in the way in which the measure of inter-cluster dissimilarity is updated after each step. The single-linkage method uses the minimum of the distances between all observations of the two sets and is indeed known to produce the \emph{subdominant ultrametric} distance of the original one \cite{mullner2011modern, rammal1986ultrametricity,carlsson2010characterization}. 
The output of this algorithm is the dendrogram shown in Fig.\ref{figSI: dendro} whose leaves correspond to each country (each 0-node) and the subdominant ultrametric distance between each pair is given by the height of the branching point of the corresponding branches.
This generates clusters of nodes that are equally  distant, in terms of $d^{\text{u}}$, one from each other. For instance, given two clusters $I$ and $J$, then the subdominant ultrametric distance between each pair of nodes belonging to $I$ and $J$ is given by $d^{\text{u}}_{i_0 j_{0}} = \min_{i_0 \in I j_{0} \in J} d_{i_0 j_0}$ for any couple $i_0 \in I, j_0 \in J$. 
Starting from the ultrametric distances $d^{\text{u}}_{i_0 j_{0}}$, we define the metaultrametric distances (related to a metametric obeying the ultrametricity condition) as
\begin{equation}
    d^{\text{mu}}_{i_0 j_{0}} =\begin{cases}
        d^{\text{u}}_{i_0 j_{0}} &\text{ if } i_0 \neq j_{0},\\
        d_{i_0 j_{0}} &\text{ if } i_0 = j_{0},\\
    \end{cases}
\end{equation}
where $d_{i_0j_0}$ is the  population averaged distance between countries $i_0$ and $j_0$ in Eq.~\eqref{Seq: Meyer dist}. 
As requested by the single linkage algorithm used to generate the dendrogram, this choice must satisfy the condition $d^{\text{mu}}_{i_0i_0} \leq \min_{j_0=1,N_0} (d^{\text{mu}}_{i_0 j_0})$. 
We found the latter inequality to hold strictly for all countries with the exception of Cameroon (which was found to be `closer' to Chad than to itself), and of Democratic Republic of the Congo and Republic of the Congo (that resulted closer to each other then to themselves). 
Given the specificity of these exceptions and the poor resulting discrepancy,in such cases we conventionally fixed $d^{\text{mu}}_{i_0 i_0} = \min_{j_0=1,N_0} (d_{i_0 j_0})$. 

\section{Network properties: empirical and expected values of the ITN} \label{SI: Net prop}
In this section, the key topological properties considered in our analysis of the directed ITN are defined.
Each such property is a function $Y(\mathbf{A}^{(\ell)})$ of the $N_\ell\times N_\ell$  adjacency matrix $\mathbf{A}^{(\ell)}$ (with entries $a^{(\ell)}_{i_\ell,j_\ell}=0,1$) of the generic $\ell$-graph. 
Note that this matrix is in general asymmetric and can contain non-zero entries along the diagonal, representing self-loops.
The matrix $\tilde{\mathbf{A}}^{(\ell)}$ represents the empirical matrix obtained at the hierarchical level $\ell$ from the Gleditsch data in year 2000 and the observed value of each considered topological property $Y$ of the ITN will be denoted as $\tilde{Y}\equiv Y(\tilde{\mathbf{A}}^{(\ell)})$.

When considering the multiscale model, $\mathbf{A}^{(\ell)}$ is instead a random matrix whose entries $\{a^{(\ell)}_{i_\ell,j_\ell}\}$ are Bernoulli random variables with expected value
\begin{equation} \label{SIeq:langle}
    \begin{split}
    \langle a^{(\ell)}_{i_\ell,j_\ell}\rangle& \equiv p_{i_\ell j_\ell} = \begin{cases}
 1- e^{-\epsilon\, \frac{\mathrm{GDP}_{i_\ell} \mathrm{GDP}_{j_\ell} }{d^{\text{mu}}_{i_\ell,j_\ell}} }&\textrm{if}\quad i_\ell \neq j_\ell\\
  1- e^{-\frac{\delta}{2}\, \frac{\mathrm{GDP}_{i_\ell}^2}{d^{\text{mu}}_{i_\ell,i_\ell}}- \Tilde{\eta} \Tilde{w}_{i_\ell}}  & \textrm{if}\quad i_\ell = j_\ell
  \end{cases}
    \end{split}
\end{equation}
where  the parameters  $(\Tilde{\eta}, \Tilde{w}_{i_0})$ are chosen so to ensure $p_{i_0 i_0}(\epsilon, \delta, \Tilde{\eta})  = 0 \quad \forall \; i_0=1,N_0 $ (in accordance with the adopted convention to fix $\tilde{a}_{i_0 i_0}^{(0)} = 0 \; \forall \, i_0$). 
As explained in the main text, this is obtained by imposing $\Tilde{\eta} =  -\frac{\delta}{2}$ and $\Tilde{w}_{i_0} = \frac{\mathrm{GDP}_{i_0}^2}{d^{\text{mu}}_{i_0 i_0}} \; \forall \, i_0 = 1,\dots,N_0$, which implies:
\begin{equation*}
    \Tilde{w}_{i_\ell} = \sum_{i_0 \in i_\ell} \frac{\mathrm{GDP}_{i_0}^2 }{d^{\text{mu}}_{i_0 i_0}} \quad \forall \ell.
\end{equation*}
Note that even though at level $\ell = 0$ both the empirical network and the multiscale model are by our choice exhibiting no self-loops, as we proceed with the coarse-graining self-loops will emerge in both the former.
This is a consequence of the coarse-graining rule $a_{i_\ell j_\ell} = 1 - \prod_{i_{\ell-1} \in  i_\ell}  \prod_{j_{\ell-1} \in  j_\ell} \left( 1- a_{i_{\ell-1} j_{\ell-1}} \right)$ and of the fact that the parameters $\Tilde{\eta} \Tilde{w}_{i_\ell}$ at aggregate levels $\ell>0$ do not nullify the first term in the exponent anymore:
\begin{equation*}
\begin{split}
    -\frac{\delta}{2} \frac{\mathrm{GDP}_{i_\ell }^2 }{d^{up}_{i_\ell i_\ell}} - \Tilde{\eta} \Tilde{w}_{i_\ell} &= -\frac{\delta}{2} \sum_{i_0 \in i_\ell}\sum_{j_0 \in i_\ell} \frac{\mathrm{GDP}_{i_0 } \mathrm{GDP}_{j_0 }}{d^{\text{mu}}_{i_0 j_0}} - \Tilde{\eta} \Tilde{w}_{i_\ell}\\
    &= \Tilde{\eta} \sum_{i_0 \in i_\ell} \sum_{j_0 \in i_\ell, j_0 \neq i_0} \frac{\mathrm{GDP}_{i_0 } \mathrm{GDP}_{j_0 }}{d^{\text{mu}}_{i_0 j_0}}\\
    &\neq 0.
\end{split}
\end{equation*}

Then, we consider the undirected projection $\mathbf{B}^{(\ell)}$ of the matrix $\mathbf{A}^{(\ell)}$, whose elements $b_{i_\ell,j_\ell}^{(\ell)} \equiv a_{i_\ell,j_\ell}^{(\ell)} + a_{j_\ell,i_\ell}^{(\ell)} - a_{i_\ell,j_\ell}^{(\ell)} a_{j_\ell,i_\ell}^{(\ell)}$ are Bernoulli random variables with expected values: 
\begin{equation} \label{SIeq:langle q}
    \begin{split}
    \langle b^{(\ell)}_{i_\ell,j_\ell}\rangle&\equiv q_{i_\ell j_\ell}  
     = \begin{cases}
  1- e^{-\delta\,  \frac{\mathrm{GDP}_{i_\ell}\mathrm{GDP}_{j_\ell}}{d^{\text{mu}}_{i_\ell,j_\ell}} } &i_\ell \neq j_\ell,\\
   1- e^{-\frac{\delta}{2}\, \frac{\mathrm{GDP}_{i_\ell}^2 }{d^{\text{mu}}_{i_\ell,i_\ell}} - \Tilde{\eta} \Tilde{w}_{i_\ell}}  &  i_\ell = j_\ell.
  \end{cases}
    \end{split}
\end{equation}

\begin{figure}[t!]
    \centering \includegraphics[width=0.4\textwidth]{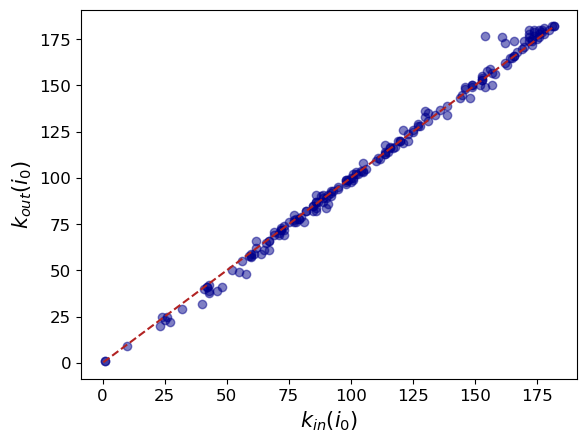}
\caption{\small\textbf{Symmetry in the WTW}: the above scatter plot illustrates the linear relation between in-degrees $\{k^{\text{in}}_{i_0}\}_{1}^{N_0}$ and out-degrees   $\{k^{\text{out}}_{i_0}\}_{1}^{N_0}$
of $0-$nodes (i.e., countries) in the World Trade Web \cite{gleditsch2002expanded}.}
    \label{fig: k_inVSk_out}
\end{figure}

The probabilities of the four possible dyads between  each pair of (block-)countries $i_{\ell}$ and $j_{\ell}$ (namely a unidirectional rightward connection, a unidirectional leftward connection, two mutual connections or a missing connection) are given by:

\begin{equation} \label{SIeq:langle freccette}
    \begin{split}
        &p_{i_\ell j_\ell }^{\rightarrow}  = \begin{cases}
        e^{-\epsilon\,  \frac{\mathrm{GDP}_{i_\ell} \mathrm{GDP}_{j_\ell} }{d^{\text{mu}}_{i_\ell,j_\ell}} }- e^{-\delta \, \frac{\mathrm{GDP}_{i_\ell} \mathrm{GDP}_{j_\ell} }{d^{\text{mu}}_{i_\ell,j_\ell}} }& i_\ell \neq j_\ell,\\
        0 & i_\ell = j_\ell,            
        \end{cases} \\
        &p_{i_\ell j_\ell}^{\leftarrow}  = \begin{cases}
        e^{-\epsilon\,  \frac{\mathrm{GDP}_{i_\ell} \mathrm{GDP}_{j_\ell} }{d^{\text{mu}}_{i_\ell,j_\ell}} }- e^{-\delta \, \frac{\mathrm{GDP}_{i_\ell} \mathrm{GDP}_{j_\ell} }{d^{\text{mu}}_{i_\ell,j_\ell}} }& i_\ell \neq j_\ell,\\
        0 & i_\ell = j_\ell,            
        \end{cases} \\
        &p_{i_\ell j_\ell }^{\leftrightarrow}= \begin{cases}
        1 - 2 e^{-\epsilon \,\frac{\mathrm{GDP}_{i_\ell} \mathrm{GDP}_{j_\ell} }{d^{\text{mu}}_{i_\ell,j_\ell}} } + e^{-\delta \,\frac{\mathrm{GDP}_{i_\ell} \mathrm{GDP}_{j_\ell} }{d^{\text{mu}}_{i_\ell,j_\ell}}} & i_\ell \neq j_\ell,\\
        1 -  e^{-\frac{\delta}{2} \,\frac{\mathrm{GDP}_{i_\ell}^2}{d^{\text{mu}}_{i_\ell,i_\ell}} - \Tilde{\eta} \Tilde{w}_{i_\ell}  } & i_\ell = j_\ell,            
        \end{cases} \\
        &p_{i_\ell j_\ell }^{\nleftrightarrow}  =\begin{cases}
        e^{-\delta\,\frac{\mathrm{GDP}_{i_\ell} \mathrm{GDP}_{j_\ell}}{d^{\text{mu}}_{i_\ell,j_\ell}}}         & i_\ell \neq j_\ell,\\
        e^{-\frac{\delta}{2} \,\frac{\mathrm{GDP}_{i_\ell}^2}{d^{\text{mu}}_{i_\ell,i_\ell}} - \Tilde{\eta} \Tilde{w}_{i_\ell}  } & i_\ell = j_\ell.            
        \end{cases}\\
    \end{split}
\end{equation}

\vspace{0.2cm}
The two remaining global parameters $\epsilon$ and $\delta$ are fitted to their unique values $\Tilde{\epsilon} \text{ and } \Tilde{\delta}$ that produce, respectively, the same number 
$L_\ell^{\text{nl}} = \sum_{i_\ell = 1}^{N_\ell} \sum_{j_\ell = 1, j_\ell \neq i_\ell}^{N_\ell} a_{i_\ell j_\ell}^{(\ell)}$ of links and  the same number 
$L_\ell^{\leftrightarrow} = \sum_{i_\ell = 1}^{N_\ell} \sum_{j_\ell = 1, j_\ell \neq i_\ell}^{N_\ell} {a_{i_\ell j_\ell}^{\leftrightarrow}}^{(\ell)}$ of reciprocated links (or, equivalently, the number 
$L_\ell^{\text{und,nl}} = \sum_{i_\ell = 1}^{N_\ell} \sum_{j_\ell < i_\ell}^{N_\ell} b_{i_\ell j_\ell}^{(\ell)} = L_\ell^{\text{nl}} - 1/2 \; L_\ell^{\leftrightarrow}$ of undirected links) as the real ITN (self-loops excluded).
In principle, this can be done at an arbitrary level $\ell$. Here, we are reporting the results obtained by constraining the above quantities at the native level $\ell = 0$:
\begin{equation} \label{Seq: contraints}
\begin{split}
    &\sum_{i_0 = 1}^{N_0} \sum_{j_0 = 1, j_0 \neq i_0}^{N_0}  p_{i_0 j_0}(\Tilde{\epsilon}) =  \Tilde{L}^{\text{nl}}_0,\\
    &\sum_{i_0 = 1}^{N_0} \sum_{j_0 = 1, j_0 \neq i_0}^{i_0} q_{i_0 j_0}(\Tilde{\delta})  = \Tilde{L}^{\text{und}, \text{nl}}_0,
\end{split}
\end{equation}
where the terms on the left side identify the expected numbers of links $\langle L_0^{\text{nl}} \rangle (\epsilon)$ and undirected links $\langle L^{\text{und},\text{nl}}_0 \rangle (\delta)$ (\emph{not} counting self-loops), while the terms on the right correspond to their empirical values. 
In the considered dataset, we find  $\Tilde{L}^{\text{nl}}_0 = 20101$ and $\Tilde{L}^{\text{und},\text{nl}}_0 = 10250$, yielding  $\Tilde{\epsilon} =  1.211 \,10^{-11}(USD)^{-2}$ and $\Tilde{\delta} = 1.344\,10^{-11}   (USD)^{-2}$.

Once the parameters $\epsilon$ and $\delta$ are fixed to $\tilde{\epsilon}$ and $\tilde{\delta}$,  we can generate realizations $\{\mathbf{A}^{(\ell)}\}$ of the $\ell$-graphs from the multiscale model at any desired hierarchical level $\ell$ by sampling $\ell$-links independently with probabilities $p_{i_\ell j_\ell}^{\rightarrow}(\tilde{\epsilon},\tilde{\delta}), p_{i_\ell j_\ell}^{\leftarrow}(\tilde{\epsilon},\tilde{\delta})$ and $p_{i_\ell j_\ell}^{\leftrightarrow}(\tilde{\epsilon},\tilde{\delta}, \tilde{\eta})$.
By averaging the value $Y(\mathbf{A}^{(\ell)})$ of a topological property over such realizations, we can efficiently estimate the corresponding expected value 
\begin{equation}
\langle{Y}\rangle\equiv\sum_{\mathbf{A}^{(\ell)}\in\mathcal{G}_{N_\ell}}P\big(\mathbf{A}^{(\ell)}\big)Y(\mathbf{A}^{(\ell)}),
\end{equation}
where $P\big(\mathbf{A}^{(\ell)} \big)$ is given by Eq.~\eqref{Deq_product2}.
If $Y(\mathbf{A}^{(\ell)})$ is linear in $\mathbf{A}^{(\ell)}$, we can  calculate $\langle{Y}\rangle$ exactly by directly replacing, in the definition of $Y(\mathbf{A}^{(\ell)})$, ${a_{i_\ell,j_\ell}^{\rightarrow}}$ with $\tilde{p}^{\rightarrow}_{i_\ell,j_\ell}$, ${a_{i_\ell,j_\ell}^{\leftarrow}}$ with $\tilde{p}^{\leftarrow}_{i_\ell,j_\ell}$ and  ${a_{i_\ell,j_\ell}^{\leftrightarrow}}$ with $\tilde{p}^{\leftrightarrow}_{i_\ell,j_\ell}$, without sampling any graph at all.
This is indeed the case for the number of total and reciprocal links in Eq.~\eqref{Seq: contraints}.

Given any $\ell$-graph $\mathbf{A}^{(\ell)}$ (be it the empirical $\ell$-graph or a random realization from the ensemble), the main topological properties of interest to us are listed below.
\begin{itemize}
    \item The \emph{link density}, representing the ratio of realized to the maximum number of links, including possible self-loops:
\begin{equation}
D_\ell(\mathbf{A}^{(\ell)})\equiv\frac{2L_\ell(\mathbf{A}^{(\ell)})}{N_\ell(N_\ell+1)}=\frac{2\sum_{i_\ell=1}^{N_\ell}\sum_{j_\ell=1}^{i_\ell}a^{(\ell)}_{i_\ell,j_\ell}}{N_\ell(N_\ell+1)};
\end{equation}
\item the \emph{reciprocity}, quantifying the proportion of mutual links with respect to the total number of links (excluding self-loops to ensure that $r_\ell(\mathbf{A}^{(\ell)}) \in [0,1]$):
\begin{equation}
r_\ell(\mathbf{A}^{(\ell)})\equiv\frac{L_\ell^{\leftrightarrow}(\mathbf{A}^{(\ell)})}{L_\ell^{nl}(\mathbf{A}^{(\ell)})}=\frac{\sum_{i_\ell=1}^{N_\ell}\sum_{j_\ell=1, j_\ell \neq i_\ell}^{N_\ell}{a^{\leftrightarrow}_{i_\ell,j_\ell}}^{(\ell)} }{\sum_{i_\ell=1}^{N_\ell}\sum_{j_\ell=1, j_\ell \neq i_\ell}^{N_\ell}{a_{i_\ell,j_\ell}}^{(\ell)}};
\label{eq:reciprocity}
\end{equation}
\item the \emph{out-degree}  (or, equivalently, the \emph{in-degree} upon exchange of the indices), counting the number of outgoing (or incoming) links from the $\ell$-node $i_\ell$:
\begin{equation}
k^{\text{out}}_{i_\ell}(\mathbf{A}^{(\ell)})\equiv\sum_{j_\ell \ne i_\ell}a^{(\ell)}_{i_\ell,j_\ell};
\end{equation}
\item the \emph{reciprocal degree}, counting the number of reciprocated links of the $\ell$-node $i_\ell$:
\begin{equation}
k^{\leftrightarrow}_{i_\ell}(\mathbf{A}^{(\ell)})\equiv\sum_{j_\ell \ne i_\ell}a^{(\ell)}_{i_\ell,j_\ell} a^{(\ell)}_{j_\ell,i_\ell};
\end{equation}
\item the \emph{undirected degree}, counting the number of connections in the undirected projection $\mathbf{B}^{(\ell)}$:
\begin{equation}
    k^{\text{und}}_{i_\ell}(\mathbf{B}^{(\ell)}) = \sum_{j_\ell \neq} b_{i_\ell j_\ell};
\end{equation}
\item the \emph{average nearest neighbour out-degree}, representing the average out-degree of the neighbours of $i_\ell$: 
\begin{equation}
{k_{\text{nn}}^{\text{out}}}_{i_\ell}(\mathbf{A}^{(\ell)})\equiv\frac{\sum_{j_\ell\ne i_\ell}\sum_{k_\ell\ne j_\ell}a^{(\ell)}_{i_\ell,j_\ell}a^{(\ell)}_{j_\ell,k_\ell}}{\sum_{j_\ell\ne i_\ell}a^{(\ell)}_{i_\ell,j_\ell}}.
\end{equation}
\end{itemize}

\begin{figure}[t!]
    \centering
    \includegraphics[width=0.4\textwidth]{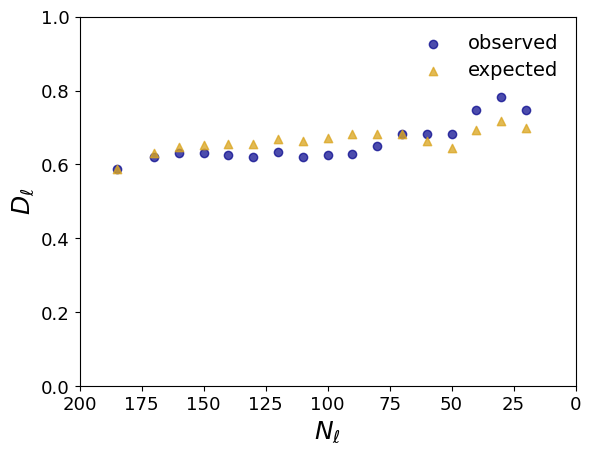} 
    \includegraphics[width=0.4\textwidth]{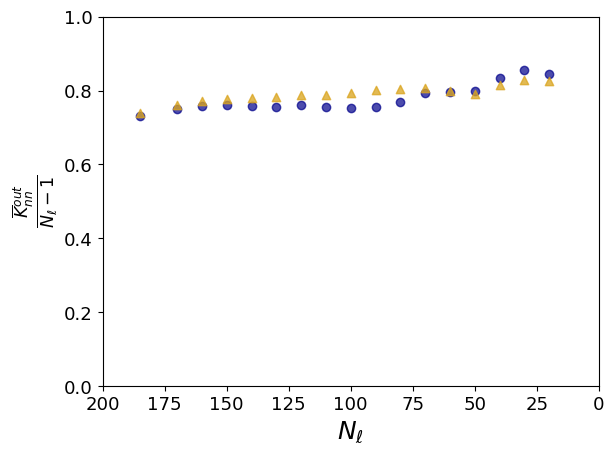}
    \caption{\small\textbf{Prediction of global topological properties of the ITN across the full spectrum of geographical aggregation using the DSIM}. The panels show the agreement between the empirical and the expected values of the link density $D_{\ell}$ (on the left) and the node-averaged rescaled average nearest neighbour out-degree  $\overline{k}_{\text{nn}}^{\text{out}}({i_{\ell}})/(N_{\ell}-1)$ as functions of the number $N_{\ell}$ of countries at all the $17$ hierarchical levels considered, with $N_0 = 185$ and $N_\ell = 180 - \ell \, 10$ for any $\ell=1,\dots 16$.}
    \label{fig:ITN-global-1}
\end{figure}

\begin{figure}[t!]
    \centering
    \includegraphics[width=0.33\textwidth]{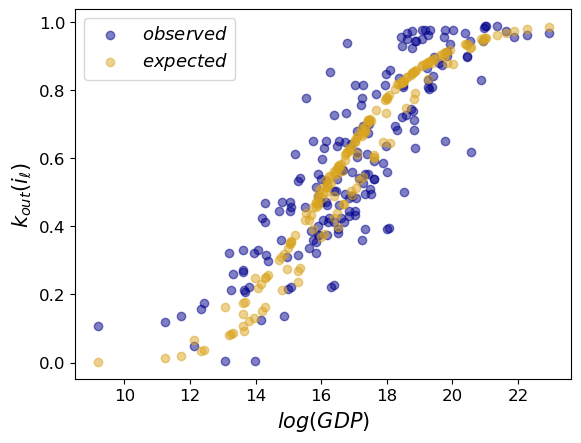}%
    \includegraphics[width=0.33\textwidth]{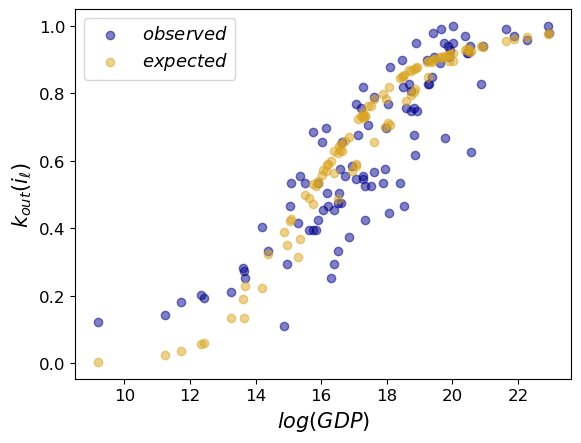}%
    \includegraphics[width=0.33\textwidth]{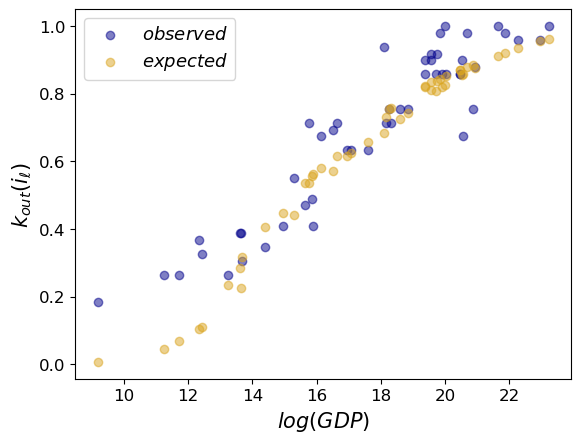}
   
    \includegraphics[width=0.33\textwidth]{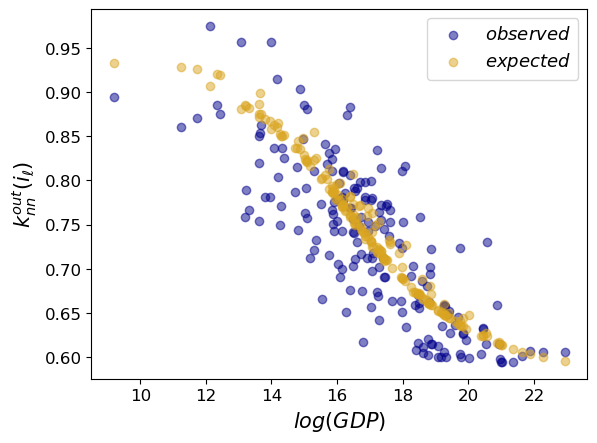}%
    \includegraphics[width=0.33\textwidth]{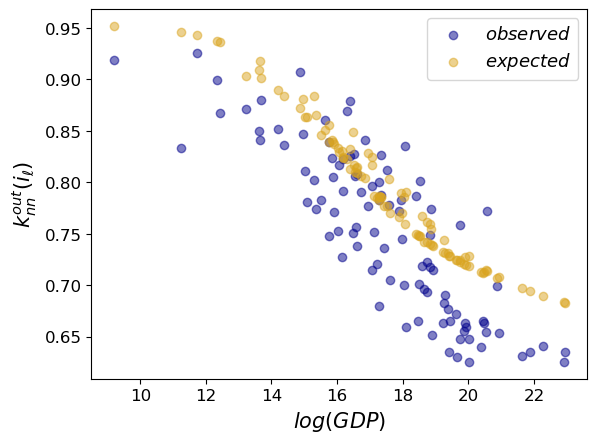}%
    \includegraphics[width=0.33\textwidth]{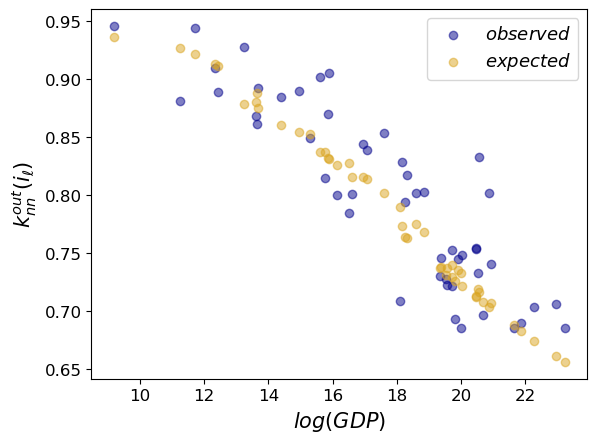}
    
    \caption{\small\textbf{Prediction of local topological properties of the ITN at different levels of aggregation using the DSIM}. Top panels: empirical (blue) and expected (yellow) out-degree $k^{out}_{i_{\ell}}$ vs $\ln(GDP_{i_{\ell}})$ for all $N_{\ell}$ nodes, for three representative hierarchical levels ($\ell_1 = 0$, $\ell_2 = 8$, $\ell_3 = 13$) such that $N_0 = 185$ (left), $N_2 = 100 $ (centre) and $N_3 = 50 $ (right). Bottom panels: empirical (blue) and expected (yellow) average nearest-neighbour out-degree $k_{\text{nn}}^{\text{out}}({i_{\ell}})$ vs $\ln(GDP_{i_{\ell}})$ for all $N_{\ell}$ nodes, for the same three hierarchical levels. }
    \label{fig:ITN-local-1}
\end{figure}

While in the main text we focused on the topological properties that are somehow affected by the extent of reciprocated links in the network ($r_\ell$ and $k_{i_\ell}$), here we note that all the quantities that were well reproduced in the undirected case, representing both global  and local properties of the network, are still well captured by the DSIM, thus confirming the desired consistency between the directed ensemble and its undirected projection. 
To illustrate this, here we consider the directed generalization of the  quantities considered in~\cite{Elenulla} and defined above: the link density $D_{\ell}$, the out-degree $k^{\text{out}}_{i_\ell}$ (the `in-' counterpart being perfectly equivalent), and the average nearest neighbour out-degree $ {k_{\text{nn}}^{\text{out}}}_{i_\ell}$. 
We performed both the global and the local comparisons and show the outcome in Figures~\ref{fig:ITN-global-1} and \ref{fig:ITN-local-1}.

\section{Network topology in the annealed scenario}
Let us start by deriving the functional form of the expected in-degree distribution. 
In close resemblance with the analysis provided in~\cite{Elenulla}, we  compute, for a typical realization of the fitness values, the distribution $P_\ell(k^{\text{in}})$  from the PDF~\eqref{eq: Levy distr} of the fitness and from the connection probability $p_{i_\ell,j_\ell}$ written as a function  of the fitness of the nodes involved:
\begin{equation}
f(x,y)=1-e^{-\epsilon\, x\,y}.
\end{equation}
For a large number $N_\ell$ of $\ell$-nodes, the expected in-degree
\begin{equation*}
    \langle k^{\text{in}}_{i_\ell} \rangle = \sum_{j_\ell \neq i_\ell} p_{i_\ell j_\ell} = \sum_{j_\ell \neq i_\ell} f(x_{i_\ell}, x_{j_\ell})
\end{equation*}
can be approximated by an integral over the number $(N_\ell-1)\varphi(y|\gamma_\ell)$ of $\ell$-nodes (except $i_\ell$ itself) with fitness in a neighbourhood of $y$. 
By denoting with $k^{\text{in}}_\ell(x)$  the expected in-degree of a node with fitness $x$ at level $\ell$, we have:
\begin{eqnarray} \label{eq: k_in(x)}
k^{\text{in}}_\ell(x) &=&(N_\ell-1)\int_0^{\infty}  f(x,y)   \varphi(y|\gamma_\ell)  \mathrm{d}y\nonumber \\ 
&=&(N_\ell-1)\left(1-e^{-\sqrt{2\epsilon \gamma_\ell x}}\right).\label{latua}
\end{eqnarray} 
The expected in-degree distribution can be exactly calculated by inverting the relation~\eqref{eq: k_in(x)} into $x(k^{\text{in}})$ and using monotonicity to write
\begin{equation}\label{Seq: continuity}
    P_\ell\left(k^{\text{in}}\right) \mathrm{d}k^{\text{in}} = \varphi\left(x(k^{\text{in}})|\gamma_\ell\right)\mathrm{d}x(k^{\text{in}}) .
\end{equation}
We therefore obtain
\begin{equation} 
    P_\ell(k^{\text{in}}) = \frac{2}{\sqrt{\pi}} \frac{\sqrt{\eta} \gamma_{\ell}}{N_\ell-1-k^{\text{in}}} \frac{e^{- \frac{\eta \gamma_{\ell}^2}{\log^2(1-\frac{k^{\text{in}}}{N_\ell-1})}}}{\log^2(1-\frac{k^{\text{in}}}{N_\ell-1})},
\end{equation}
as shown in the main text.

Similarly, to evaluate the reciprocated degree distribution $\Tilde{P}_\ell(k^{\leftrightarrow})$, we  consider the function:
\begin{equation}
f^{\leftrightarrow}(x,y)=1+e^{-\delta\, x\,y}-2e^{-\eta\, x\,y}.
\end{equation}
The expected reciprocal degree $k_\ell^{\leftrightarrow}$ of a node with fitness $x$, for a large number $N_\ell$ of nodes, is given by:
\begin{eqnarray} 
k^{\leftrightarrow}_\ell(x) &=&(N_\ell-1)\int_0^{\infty}  f^{\leftrightarrow}(x,y)   \varphi(y|\gamma_\ell)  \mathrm{d}y\nonumber \\ 
&=&(N_\ell-1)\left(1-2 e^{-\sqrt{2\epsilon \gamma_\ell x}} + e^{-\sqrt{2\delta \gamma_\ell x}}\right).\label{elasua}
\end{eqnarray} 
As pointed out in the main text, the above expression  cannot be inverted in general, and thus we cannot rely on Eq.~\eqref{Seq: continuity} for evaluating the distribution $\Tilde{P}_\ell(k^{\leftrightarrow})$.
Nevertheless, we can consider an approximation for small values of $x$, in which case $k_{\ell}^{\leftrightarrow}(x) \sim (N_\ell -1) \sqrt{2 \gamma_{\ell} x} \;(2 \sqrt{\epsilon} - \sqrt{\delta})$. 
The corresponding inverse function $x(k_{\ell}^{\leftrightarrow})$ can easily be found:
\begin{equation*} \label{Seq: k_rec VS x}
    x(k_{\ell}^{\leftrightarrow}) \approx \frac{1}{2 \gamma_\ell} \left(\frac{k}{(N_\ell -1)(2\sqrt{\eta} - \sqrt{\delta})} \right)^2.
\end{equation*}
Although the above function is not monotonic in $k$, we can still use Eq.~\eqref{Seq: continuity} in our domain of interest, which is the positive real axis.
The (approximated) reciprocal degree distribution therefore is given by:
\begin{equation}\label{Seq: cumulative rec-deg}
\Tilde{P}_\ell(k^{\leftrightarrow}) \approx \frac{2}{\sqrt{\pi}} (N_\ell -1) \frac{\gamma_\ell \left(2\sqrt{\eta} - \sqrt{\delta}\right)}{\left(k^{\leftrightarrow}\right)^2}  e^{-\left((N_\ell -1) 
    \frac{\gamma_{\ell}(2 \sqrt{\eta} - \sqrt{\delta})}{k^{\leftrightarrow}}\right)^2} .
\end{equation}

\end{document}